\documentclass[aps,prc,reprint,superscriptaddress,showpacs,nofootinbib]{revtex4-1}

\usepackage{color}
\usepackage{framed}
\usepackage{amsmath}
\usepackage{amssymb}
\usepackage{graphicx}
\usepackage{bm}
\usepackage[normalem]{ulem}

\def\del{\partial}
\def\dis{\displaystyle}
\def\la{\langle}
\def\ra{\rangle}
\def\l|{\left|}
\def\r|{\right|}
\def\<{\left<}
\def\>{\right>}

\newcommand{\be}{\begin{eqnarray}}
\newcommand{\ee}{\end{eqnarray}}
\newcommand{\ket}{\rangle}
\newcommand{\bra}{\langle}
\newcommand{\calL}{{\cal L}}

\begin{document}

\title{
Structure of charmed baryons studied by pionic decays
} 

\author{Hideko Nagahiro}
\thanks{}
\affiliation{Department of Physics, Nara Women's University, 
Nara 630-8506, Japan}
\affiliation{Research Center for Nuclear Physics (RCNP), Osaka
University, Ibaraki, Osaka 567-0047, Japan}

\author{Shigehiro Yasui}
\affiliation{Department of Physics, Tokyo Institute of Technology,
Meguro 152-8551, Japan}

\author{Atsushi Hosaka}
\affiliation{Research Center for Nuclear Physics (RCNP), Osaka
University, Ibaraki, Osaka 567-0047, Japan}
\affiliation{J-PARC Branch, KEK Theory Center, 
KEK, Tokai, Ibaraki 319-1106, Japan}

\author{Makoto Oka}
\affiliation{Department of Physics, Tokyo Institute of Technology,
Meguro 152-8551, Japan} 
\affiliation{Advanced Science Research Center, Japan Atomic Energy
Agency, Tokai, Ibaraki, 319-1195, Japan}

\author{Hiroyuki Noumi}
\affiliation{Research Center for Nuclear Physics (RCNP), Osaka
University, Ibaraki, Osaka 567-0047, Japan}

\date{\today}

\begin{abstract}
We investigate the decays of the charmed baryons aiming at the
 systematic understanding of hadron {internal} structures based on the
 quark model by paying attention to heavy quark symmetry. We evaluate the decay 
 widths from the one pion emission 
 for the {known} excited states, $\Lambda_c^*(2595)$, $\Lambda_c^*(2625)$,
 $\Lambda_c^*(2765)$, 
 $\Lambda_c^*(2880)$ and $\Lambda_c^*(2940)$, as well as for the ground
 states $\Sigma_c(2455)$
 and $\Sigma_c^*(2520)$. 
{The decay properties of the lower excited charmed baryons are well
 explained, and several important predictions for 
higher excited baryons are given.}
 We find that the
 axial-vector type coupling of the pion to the light quarks is
 essential,
{which is expected from chiral symmetry,}
 to reproduce the decay widths especially of the low lying
 $\Lambda_c^*$ baryons.
 We emphasize the {importance} of the branching ratios of
 $\Gamma(\Sigma_c^*\pi)/\Gamma(\Sigma_c\pi)$ for the study of the
 nature of higher excited  $\Lambda_c^*$ baryons.

\end{abstract}


\maketitle

%
\section{Introduction}

{Understanding of the internal structure of hadrons 
is an important 
subject in hadron physics}.
One of the most important problems is to identify the effective
degrees of freedom which should play essential roles at low energies,
because the bare quarks do not appear at such 
a scale due to the color confinement of QCD.
To identify the effective degrees of freedom should
serve not only for the understanding of the QCD vacuum properties,
but also be useful to explain and predict experimental data with simple
physical terms. 
%
%
%
In this respect, what we are
aiming at is to establish the economized effective 
{degrees of freedom}
for {various phenomena of the} strong interaction
physics~\cite{Weinberg:1962hj,*Weinberg:1965zz,Nagahiro:2014mba}.

The charmed baryons, containing a single heavy charm quark, 
is a good place to study the hadron
structure.
One of the important features is the spin symmetry of the heavy quark.
QCD predicts that the spin-dependent interaction of the heavy quark is
suppressed by $1/m_Q$ and thus in the infinite $m_Q$ limit, the heavy
quark spin is decoupled from the dynamics of the light quarks. 
%
{The dynamical decoupling of the light quark spin and the heavy
quark spin} is the heavy quark symmetry
{(HQS)}~\cite{Isgur:1991wq}.   

{In the heavy quark limit, the light quark component is called the
{\it brown muck} as the colorful object conserving its total
spin~\cite{Neubert:1993mb}. 
In terms of QCD, the brown muck contains not only
light quarks but also light antiquarks as well as gluons. 
For the spin of the brown muck $j$, the heavy hadrons are classified to
one state with the total spin $J=1/2$ for $j=0$ and two degenerate states
with the total spin $J=j\pm1/2$ for $j\ge1/2$. 
The former is called the HQS singlet, and the latter is called the HQS doublet.
The classification based on the HQS is useful for the investigation of
the heavy hadrons, because the brown muck spin serves as an additional conserved
quantum number reflecting the internal structure of the heavy
hadrons.} 
The HQS appears in many properties of heavy hadrons,
such as the mass spectrum and the decay branching ratios\footnote{The
heavy quark symmetry can be applied also to exotic heavy hadrons such as
hadronic molecules~
\cite{Yasui:2009bz,Yamaguchi:2011qw,Yamaguchi:2011xb,Yamaguchi:2013hsa,GarciaRecio:2008dp,Gamermann:2010zz,GarciaRecio:2012db,Yamaguchi:2014era,Liu:2011xc,Maeda:2015hxa,Yasui:2013vca} 
as well as to the heavy hadrons in nuclear
medium~\cite{Yasui:2013vca,Yasui:2012rw,Yasui:2013iga,Yasui:2014cwa,Suenaga:2014dia,Suenaga:2014sga,Suenaga:2015daa}. See Ref.~\cite{Hosaka:2016ypm}
as a review for the latter.}.

%
%
%

{There is another interesting feature of the charmed baryons.
In the quark model description, {we have} two
different orbital {motions in the low energy} excitations.}
One 
is {the} relative motion between two light quarks, so-called
{$\rho$}-mode.
{T}he other {is the one}
between 
the center-of-mass of the two light quarks {and the charm quark},
so-called {$\lambda$}-mode. 
 Owing to the mass
difference of the {light and heavy} quarks, the excitation energies {of} the
$\lambda$- and $\rho$-modes are kinematically well separated, and the
internal excitations are dominated
{exclusively} by either {$\rho$-mode or $\lambda$-mode}
with only small mixing~\cite{Yoshida:2015tia}.
This contrasts with light quark baryons where the two modes generally
mix largely, and thus is the reason that we can study the two basic
modes exclusively in the heavy baryons. 

{In general, internal structures of hadron are 
reflected not only in mass spectrum but also in various transition
properties such as productions and decays.
Among them, two-body decay processes through the {one-}pion emission are
particularly interesting due to the following reasons: 
(i) The pion couples only to the light quarks,
and the charm quark behaves simply as a spectator.
The dynamics of the pion
is governed by chiral symmetry in a unique manner.  Therefore, {the}
transitions accompanying pion emission should bring important information
about the dynamics of the two light quarks in a heavy baryon.  This
is also helpful to understand 
diquark properties in a heavy baryon.
(ii) Some low-lying states of excited charmed baryons have significantly
smaller excitation energies than light baryon excitations, and the
emitted pion carries only a small momentum.  
{Therefore, the pion emission from the excited charmed baryons is a good
place to study the quark-pion} interaction, which
{should be} well determined by the
low energy chiral dynamics. {This} can be checked by {comparing the
theoretical results with} the {observed} decays of 
the low-lying charmed baryons.}

\begin{figure}[thb]
 \includegraphics[width=\linewidth]{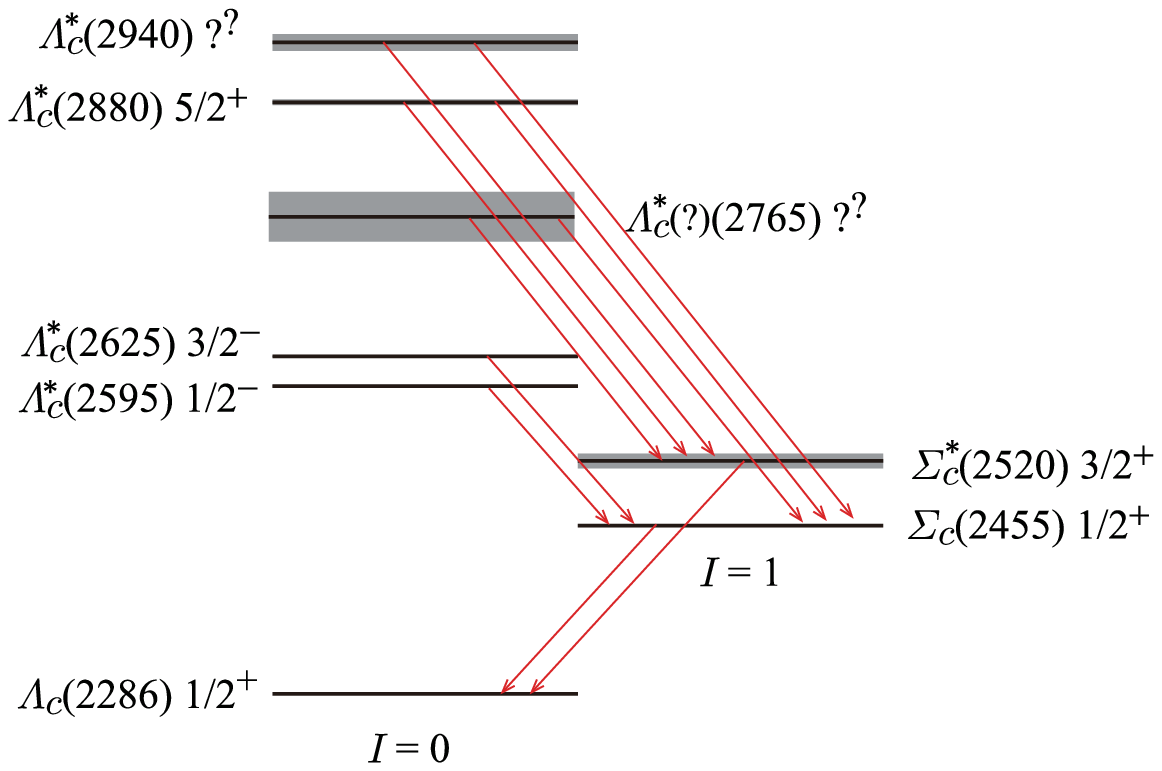}
\caption{(color online) Level structure of
 the charmed baryons with {isospin} $I=0$ and $I=1$ $Y_c(mass) J^P$
 {considered in this study.} 
The hatched squares denote their total decay widths in
 Particle Data Group (PDG)~\cite{Agashe:2014kda}.
The arrows indicate the possible decay paths with one-pion emission evaluated
 in this study.
\label{fig:level}}
\end{figure}

{In} this paper we consider the pion emission
decays from the orbitaly excited charmed baryons~\footnote{{In this
article, we express the ground and excited charmed baryons as $Y_c$ and $Y_c^*$.}}
$\Lambda_c^*(2595)$,
$\Lambda_c^*(2625)$, 
$\Lambda_c^*(2765)$,
$\Lambda_c^*(2880)$,
$\Lambda_c^*(2940)$ into $\Sigma_c(2455)\pi$ and $\Sigma_c^*(2520)\pi$,
and those from
{orbital} ground state charmed baryons 
$\Sigma_c(2455)$ and $\Sigma_c^*(2520)$ into $\Lambda_c(2286)\pi$.
{The {decay paths} are summarized in Fig.~\ref{fig:level}.  
To estimate the decay widths numerically,
we employ a non-relativistic constituent quark model with a harmonic
oscillator {potential as the quark confinement force}.    
{The model is rather simple but we expect that 
essential and universal features can be extracted.}

{There are previous works investigating strong decays of charmed
baryons~\cite{Isgur:1991wq,
Yan:1992gz,
Cho:1992gg,
Cho:1994vg,
Rosner:1995yu,
Albertus:2005zy,
Cheng:2006dk,
Zhong:2007gp,
Yasui:2014cwa}.  
In Ref.~\cite{Cheng:2006dk}, based on heavy hadron chiral perturbation theory 
the importance of heavy quark symmetry are discussed
{in the heavy quark limit}. 
{In Ref.~\cite{Yasui:2014cwa}, including the correction terms from the
next-to-leading order ${\cal O}(1/m_{Q})$, relationships between decay
widths in several decay channels were obtained.} 
In Ref.~\cite{Zhong:2007gp}, non-relativistic quark model calculations were
performed and decays of various quark model states were investigated.
In the present study, we will also employ the non-relativistic quark model.
It is worthwhile to emphasize the difference between the works in
Ref.~\cite{Zhong:2007gp} and ours.
In Ref.~\cite{Zhong:2007gp}, the 
baryon wave functions are constructed in the so-called $LS$ coupling
scheme, while we do in the $jj$ coupling scheme where the brown muck
total $j$ is first formed.   
In doing so, we will derive various relations and selection rules
{in relation to} HQS.

In our study, we will shed light upon the following issues.
Firstly, we check the validity of the present framework by
calculating the decay widths of the two $\Sigma_c$ baryons,
$\Sigma_c(2455)(J^P=1/2^+)$ and 
$\Sigma_c^*(2520)(J^P=3/2^+)$, which are 
the {orbital} ground state of charmed baryons.
These baryons decay into $\Lambda_c(2286)\pi$ 
{as} the only possible  channel in strong decay.  Because both
the initial and final charmed baryon states are in the {orbital} ground states
in the quark model, those charmed baryons are good objects for
confirmation of the validity of our formalism 
for the one-pion emission.
We will see that our results
are in reasonably good agreement with the experimental values.

{Secondly, we investigate the decay properties of}
the $\Lambda_c^*(2595) (J^P=1/2^-)$ and $\Lambda_c^*(2620) (J^P=3/2^-)$
as
the lowest-lying orbital excitations in 
$p$-wave.
{They are interesting because they have the subcomponent,} the spin-0
diquark system, which is {moving in the $p$-wave orbital of} {the} 
$\lambda$-mode~\cite{PhysRevD.20.768,Cho:1994vg,Pirjol:1997nh}.  
{They}
have been {observed} in 
$e^+e^-$ {collisions} and
$p\bar{p}$ {collisions}~\cite{Albrecht:1993pt,Edwards:1994ar,Aaltonen:2011sf}
as well as in photoproductions~\cite{Frabetti:1993hg}.
{An interesting feature of them is} that the
$\Lambda_c^*(2595)(1/2^-)$ baryon has a considerably large decay width
into $\Sigma_c\pi$ channel although its phase space is 
very small.  In contrast, $\Lambda_c^*(2625)(3/2^-)$ has a very small width
{although there is}
sufficiently large phase space {in its decay channel $\Sigma_c\pi$}.  We show
that the quark model description with the $\lambda$-mode can explain
these decay properties  
very well {for these low-lying $Y_c$ states}.
We find that, to achieve the good agreements, 
the $\pi qq$ interaction
Lagrangian 
of the derivative
coupling (axial-vector coupling) is needed to reproduce the experimental
decay width.
This strongly implies that the non-linear chiral
dynamics works for the pion and constituent quarks.
{We will present that} especially
decay properties of $\Lambda_c^*(2595)$ are much affected by the isospin
breaking effect near the thresholds. 

{Thirdly,}
we {study} higher excited charmed baryons, 
$\Lambda_c^*(2765)$, $\Lambda_c^*(2880)$ and $\Lambda_c^*(2940)$.
Because their spins and parities are not fully determined
experimentally, we consider various {patterns of} assignments of $1/2^\pm$, $3/2^\pm$
and $5/2^\pm$ which are formed by the quark model.
By comparing the resulting decay widths with existing experimental data,
we will see that several assignments of spin and parity will be excluded.

{Finally, we will pay special attention to $\Lambda_c^*(2880)$ for
the determination of its spin and parity.}
In PDG~\cite{Agashe:2014kda}, the spin of the
$\Lambda_c^*(2880)$ is $5/2$ which is {determined by}
the angular distribution of
$\Sigma_c(2455)\pi$ decay~\cite{Abe:2006rz,Agashe:2014kda}, and the
{positive} parity is {inferred from the}
agreement of the {observed} decay branching ratio $\Sigma_c^*(2520)/\Sigma_c(2455)$ 
{in comparison with the prediction from} heavy quark
symmetry~\cite{Abe:2006rz,Isgur:1991wq,Cheng:2006dk}.  As carefully
argued in Ref.~\cite{Cheng:2006dk}, however, possible $p$-wave contribution {was} simply ignored {in the evaluation of the
branching ratio}.
We show that the {many} configurations for the $\Lambda_c^*$
baryons with $J^P=5/2^+$ are turned out to be incompatible with the
{present} experimental 
data~\cite{Abe:2006rz} if the $p$-wave {contribution} is {properly} considered.
We find that only one configuration leads to the result consistent
with the data where $p$-wave contribution vanishes due to the selection
rule working for the pion emission between diquarks,
the occurrence of which is a unique feature of heavy baryons where a
heavy quark behaves as a spectator, namely in heavy quark symmetry.

{This article is organized as follows.
In Sec.~\ref{sec:WF}, we {explain} wave functions of the charmed baryons
employed in {our} constituent quark model.  In
Sec.~\ref{sec:formalism} we present {the} formalism for the one-pion
emission decay of the charmed baryon.  We show our numerical results for
the decay widths in Sec.~\ref{sec:results}.  Finally,
Sec.~\ref{sec:summary} is devoted to the summery.}

\section{Baryon wave functions within the quark model}
\label{sec:WF}

We construct the baryon wave functions in a scheme
   inspired by the heavy quark symmetry.  Namely, first we construct a
   brown muck wave function using light degrees of freedom, which is
   then combined with the heavy quark to form the total baryon wave
   functions.  In this manner, we will be able to see in a transparent
   manner various relations and selection rules which are valid in the
   heavy quark limit.   
   Let us start with the harmonic oscillator Hamiltonian for the orbital
   wave function,
\begin{equation}
 H={-}\sum_{i=1}^3\frac{\vec{\nabla}_i^2}{2m_i} + \sum_{i\ne
  j}\frac{k}{2}(\vec{r}_i-\vec{r}_j)^2,
\label{eq:H}
\end{equation}
where $\vec{r}_i$ are the spatial coordinate{s} of the
{\it i}-th quark of mass $m_i$ and $k$ the spring constant.  

{Quark-1 and quark-2 denote the two light quarks of mass $m$
($m_1=m_2=m$), and quark-3 {the} charm quark of mass $M$,
$(m_3=M)$.} 
The Hamiltonian can be divided {into} {one} for the
center-of-mass motion 
{$\vec{X}$} and {those for the relative}
motions {$\vec{\rho}$ and
$\vec{\lambda}$} as 
\begin{equation}
 H=H_G+H_\rho+H_\lambda,
\end{equation}
where 
\begin{subequations} 
\begin{eqnarray}
 H_G &=& {-}\frac{\vec{\nabla}_X^2}{2(2m+M)}\ ,\\
 H_\rho&=& {-}\frac{\vec{\nabla}_\rho^2}{2m_\rho}+\frac{m_\rho
  \omega_\rho^2}{2}\vec{\rho}^{\,2} \ ,\\
 H_\lambda&=& {-}\frac{\vec{\nabla}_\lambda^2}{2m_\lambda}+\frac{m_\lambda
  \omega_\lambda^2}{2}\vec{\lambda}^{\,2} \ .
\end{eqnarray}
\end{subequations} 
Here, the coordinate of the center-of-mass {$\vec{X}$ is {defined as}}
\begin{equation}
  \vec{X}=\frac{1}{2m+M}(m(\vec{r}_1+\vec{r}_2)+M\vec{r}_3),
\end{equation}
and $\vec{\rho}$ and $\vec{\lambda}$ are the Jacobi coordinates
defined as
\begin{subequations} 
\begin{eqnarray}
\vec{\rho}&=&\vec{r}_1-\vec{r}_2, \\
 \vec{\lambda}&=&\frac{1}{2}(\vec{r}_1+\vec{r}_2)-\vec{r}_{3}.
\end{eqnarray}
\end{subequations}
\begin{figure}[hbt]
 \includegraphics[width=0.25\linewidth]{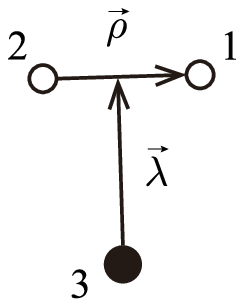}
\caption{Difinitions of the Jacobi coordinates $\vec{\rho}$ and
 $\vec{\lambda}$.  The quarks 1 and 2 are the light quarks, and 3
 the heavy (charm) quark.
\label{fig:rho_lambda}} 
\end{figure}
{As indicated in Fig.~\ref{fig:rho_lambda}, $\vec{\rho}$ is}
the relative coordinate between {the} two light quarks (quark-1 and
quark-2), and {$\vec{\lambda}$ is} the relative coordinate between
the center-of-mass of the two light quarks and the charm
quark. 

The reduced masses $m_\lambda$ {and} $m_\rho$ are defined by
\begin{equation}
 m_\rho=\frac{m}{2}, \ \ m_\lambda=\frac{2mM}{2m+M},
\label{eq:reducedmass}
\end{equation}
and the frequencies of the oscillator for $\lambda$- and $\rho$-modes are by
\begin{equation}
 \omega_\rho=\sqrt{\frac{3k}{m}},\ \ 
 \omega_\lambda=\sqrt{\frac{k(2m+M)}{mM}}.
\label{eq:omega}
\end{equation}

Orbital wave functions of the three-quark state are expressed by a
simple product of the eigenfunctions of the {separated} Hamiltonians
\begin{equation}
 \Psi(\vec{r}_1,\vec{r}_2,\vec{r}_3)=\psi{_{\lambda}}(\vec{\lambda})\psi{_{\rho}}(\vec{\rho})e^{i\vec{P}\cdot\vec{X}}
  \ , 
\end{equation}
where $\vec{P}$ is {the} total momentum of the three-quark state, and
$\psi{_{\lambda}}(\vec{\lambda})$ and $\psi{_{\rho}}(\vec{\rho})$ the wave
functions of the Jacobi coordinates {$\vec{\lambda}$} and {$\vec{\rho}$}.
The
wave function{s} of the harmonic oscillator {are} given by
\begin{equation}
 \psi_{n\ell m}(\vec{x})=R_{n\ell}(r)Y_{\ell m}(\hat{x}) \ ,
\label{eq:ws}
\end{equation}
 where the radial function $R_{n\ell}(r)$ is summarized in Appendix {and $Y_{\ell m}$ is the spherical harmonics}.
{We will call the excitation with either $n_{\lambda}\neq0$ (radial excitation) or $\ell_{\lambda}\neq0$ (orbital excitation) the $\lambda$-mode.
This is also the case for the $\rho$-mode.
When both $\lambda$-mode and $\rho$-mode happen, this is called the $\lambda\rho$-mode.
}

The full wave functions of baryons {are} constructed by {products} {of}
isospin (flavor) {part, spin part, and the orbital part.  For the isospin
part, we introduce the notation $D^{I}_{(I_{z})}$ for the two light quarks as
\begin{equation}
D^0:\left\{D^0_0 = \frac{1}{\sqrt{2}}(ud-du)\right\}  \ ,
\end{equation}
for $I=0$ state, and
\begin{equation}
D^1:\left\{ D^1_1=uu,\ \ D^1_0=\frac{1}{\sqrt{2}}(ud-du), \ \ D^1_{-1}=dd\right\},
\end{equation}
for $I=1$ state{s}.  The flavor wave function of the $\Lambda_c$ baryons
having $I=0$ is then expressed by $D^0c$ ($c$ stands for the charm quark),
and that of the $\Sigma_c$ baryons with $I=1$ is by $D^1c$.

Similarly, the spin wave functions of the two light quarks {are} expressed
by $d^{s}_{(s_{z})}$,
\begin{gather}
 d^0:\left\{d^0_0=\frac{1}{\sqrt{2}}(\uparrow \downarrow - \downarrow
 \uparrow) \right\}\ , \\
d^1:\left\{d^1_1= \uparrow \uparrow, \ 
d^1_0= \frac{1}{\sqrt{2}}(\uparrow \downarrow + \downarrow
 \uparrow), \ 
d^1_{-1}=\downarrow\downarrow \right\}\ .
\end{gather}
For the charm quark spin, we use the symbol $\chi_c$ {for either spin up or down}.  

By making use of these expressions, the full wave functions of the
$\Lambda_c(J)$ and $\Sigma_c(J)$ {with total spin $J$} {are} constructed as
\begin{gather}
 \Lambda_c(JM) = \left[
[\psi_{{n_\lambda\ell_\lambda
 m_\lambda}}(\vec{\lambda})\psi_{{n_\rho\ell_\rho m_\rho}}(\vec{\rho}),
d]^{j},
\chi_c\right]^{J}_M D^0c  \ ,\\
\Sigma_c(JM)=
\left[
[\psi_{{n_\lambda\ell_\lambda m_\lambda}}(\vec{\lambda})\psi_{{n_\rho\ell_\rho m_\rho}}(\vec{\rho}),
d]^{j},
\chi_c\right]^{J}_M D^1c \ ,
\end{gather} 
by anti-symmetrizing {the light quark part} including the color
part which is not explicitly shown 
here. The total spin $J$ of the charmed baryon is given by the sum of
the spin of 
charm quark and the ``total'' angular momentum $j$ of all the remaining part
(so-called {\em brown muck} \cite{Neubert:1993mb}) 
which is obtained by composing the orbital angular momenta $\ell_\lambda$
and $\ell_\rho$ and diquark spin $d$.
For example, the {wave functions of orbital} ground state
for the charmed baryons are given by
\begin{gather}
 \Lambda_c(1/2^+)=
\left
[
[\psi_{{0s}}(\vec{\lambda})\psi_{{0s}}(\vec{\rho}),
d^0]^{0},
\chi_c\right 
]^{1/2} D^0c\ ,\\
 \Sigma_c(1/2^+)=
\left[
[\psi_{{0s}}(\vec{\lambda})\psi_{{0s}}(\vec{\rho}),
d^1]^{1},
\chi_c\right]^{1/2} D^1c \ ,
\label{eq:Sigma_gs}
\end{gather} 
and 
\begin{equation}
 \Sigma_c^*(3/2^+)=
\left[
[\psi_{{0s}}(\vec{\lambda})\psi_{{0s}}(\vec{\rho}),
d^1]^{1},
\chi_c\right]^{3/2} D^1c \ .
\end{equation}


\begin{table}[htp]
\caption{Quark configurations considered in this
 article. $(n_{\lambda(\rho)},\ell_{\lambda(\rho)})$ are 
the {nodal} and the
angular momentum {quantum numbers} for the $\lambda(\rho)$ motion
 wave function.  The spin {wave function} of the two light quarks
 is expressed by {$d$}.
 The brown muck spin and the parity is expressed by $j^P$. The total angular
 momentum $\vec{\ell}=\vec{\ell}_\lambda+\vec{\ell}_\rho$ are also shown 
{for $\lambda\rho$-mode.}
The spin
 and party $J^P$ and supposed physical charmed baryons are also shown.
\label{tab:config}}
\begin{center}
\begin{tabular}{ccccc|c} 
\multicolumn{6}{l}{Ground states charmed baryons} \\\hline\hline
$(n_\lambda,\ell_\lambda)$ & $(n_\rho,\ell_\rho)$ & $d{^{s}}$ & $j^P$ & $J^P$ & {possible assignment} \\\hline\hline
$(0,0)$ & $(0,0)$ & $d^0$ & $0^+$ & $1/2^+$ & $\Lambda_c(2286)$ \\
$(0,0)$ & $(0,0)$ & $d^1$ & $1^+$ & $(1/2, 3/2)^+$ & $\Sigma_c(2455)$, $\Sigma_c^*(2520)$ \\
 &  &  &  &  &  \\
\multicolumn{6}{l}{Negative parity excited charmed baryons} \\\hline\hline
$(n_\lambda,\ell_\lambda)$ & $(n_\rho,\ell_\rho)$ & $d{^{s}}$ & $j^P$ & $J^P$ & {possible assignment} \\\hline\hline
$(0,1)$ & $(0,0)$ & $d^0$ & $1^-$ & $(1/2, 3/2)^-$ & $\Lambda_c^*(2595)$, $\Lambda_c^*(2625)$ \\
$(0,0)$ & $(0,1)$ & $d^1$ & $0^-$ & $1/2^-$ &  \\
 &  &  & $1^-$ & $(1/2, 3/2)^-$ &  \\
 &  &  & $2^-$ & $(3/2, 5/2)^-$ & $\Lambda_c^*(2880)$(?) \\
 &  &  &  & \multicolumn{1}{c}{} & \multicolumn{1}{c}{} \\
\multicolumn{6}{l}{Positive parity excited charmed baryons} \\\hline\hline
$(n_\lambda,\ell_\lambda)$ & $(n_\rho,\ell_\rho)$ & $d{^{s}}$ & $j^P$ & $J^P$ & {possible assignment} \\\hline\hline
$(1,0)$ & $(0,0)$ & $d^0$ & $0^+$ & $1/2^+$ &  \\
$(0,2)$ & $(0,0)$ & $d^0$ & $2^+$ & $(3/2, 5/2)^+$ & $\Lambda_c^*(2880)$(?) \\
$(0,0)$ & $(1,0)$ & $d^0$ & $0^+$ & $1/2^+$ &  \\
$(0,0)$ & $(0,2)$ & $d^0$ & $2^+$ & $(3/2, 5/2)^+$ & $\Lambda_c^*(2880)$(?) \\
 &  &  &  &  &  
\end{tabular}
\begin{tabular}{cccccc|c} 
\multicolumn{7}{l}{Positive parity excited charmed baryons ({$\lambda\rho$}-mode)} \\\hline\hline
$(n_\lambda,\ell_\lambda)$ & $(n_\rho,\ell_\rho)$ & $d{^{s}}$ & {$\ell$} & $j^P$ & $J^P$ & {possible assignment}  \\\hline\hline
$(0,1)$ & $(0,1)$ & $d^1$ & 0 & $1^+$ & $(1/2, 3/2)^+$ &  \\\cline{4-7}
 &  &  & 1 & $0^+$ & $1/2^+$ &  \\
 &  &  &  & $1^+$ & $(1/2, 3/2)^+$ &  \\
 &  &  &  & $2^+$ & $(3/2, 5/2)^+$ & $\Lambda_c^*(2880)$(?) \\\cline{4-7}
 &  &  & 2 & $1^+$ & $(1/2, 3/2)^+$ &  \\
 &  &  &  & $2^+$ & $(3/2, 5/2)^+$ & $\Lambda_c^*(2880)$(?) \\
 &  &  &  & $3^+$ & $(5/2, 7/2)^+$ & $\Lambda_c^*(2880)$(?) \\
\end{tabular}
\end{center}
\end{table}

{In Table \ref{tab:config}, we summarize the quark configurations for
the charmed baryons considered in this article.}
The observed $\Lambda_c$ excited states $\Lambda_c^*(2595)$ and
$\Lambda_c^*(2625)$ baryons are, {due to their small excitation
energies, assigned to be the $p$-wave excitations of the $\lambda$-mode
({$n_{\lambda}=0$,} $\ell_\lambda=1$) with spin-0 diquark {($d^{0}$)}.}
Their {quark} configurations are given by
\begin{eqnarray}
  \Lambda_c^*(1/2^-;\lambda\text{-mode})=
\left[
[\psi_{{0p}}(\vec{\lambda})\psi_{{0s}}(\vec{\rho}),
d^0]^{1},
\chi_c\right]^{1/2} D^0c \ , \nonumber \\
\end{eqnarray}
and
\begin{eqnarray}
  \Lambda_c^*(3/2^-;\lambda\text{-mode})=
\left[
[\psi_{{0p}}(\vec{\lambda})\psi_{{0s}}(\vec{\rho}),
d^0]^{1},
\chi_c\right]^{3/2} D^0c \ . \nonumber \\
\end{eqnarray}

Another possibility to construct the negative parity excited
states for $\Lambda_c^*$ is the so-called $\rho$-mode excitation 
({$n_{\rho}=0$,} $\ell_\rho=1$),  which must have the spin-1 diquark {($d^{1}$)}
due to the anti-symmetrization of the wave function. The total
spin $j$ of the brown muck can be $j=0$, 1 and 2,  {leading} to a {HQS} singlet
with the baryon spin $J=1/2$, and two {HQS} doublets $J=(1/2,3/2)$ and
$J=(3/2,5/2)$, respectively.  {For example,} the concrete form {for the
HQS singlet} is given by 
\begin{eqnarray}
 \Lambda_c^*(J^-;\rho\text{-mode})=
\left[
[\psi_{{0s}}(\vec{\lambda})\psi_{{0p}}(\vec{\rho}),
d^1]^{j},
\chi_c\right]^{J=j\pm 1/2} D^0c \ . \nonumber \\
\end{eqnarray}


The minimal configuration for $J^P=1/2^+$ state {for $\Lambda_c$
baryons} is an orbital excitation
for the {nodal} quantum number $n{_{\lambda}}=1$ {or $n_{\rho}=1$} as {with spin-0 diquark given by}
\begin{eqnarray}
\Lambda_c^*(1/2^+;n_\lambda\!=\!1)&=&
\left[
[\psi_{{1s}}(\vec{\lambda})\psi_{{0s}}(\vec{\rho}),
d^0]^{0},\chi_c
\right]^{1/2}. \\
  \Lambda_c^*(1/2^+;n_\rho\!=\!1)&=&
\left[
[\psi_{{0s}}(\vec{\lambda})\psi_{{1s}}(\vec{\rho}),
d^0]^{0},\chi_c
\right]^{1/2},
\end{eqnarray}
both of which are the HQS singlets.

{The higher excited states of {$J^{P}$ with $P=+$} can be constructed by the
 $d$-wave excitation  {as the total angular momentum}. In this case, w}e
have three 
possibilities as 
$(\ell_\lambda,\ell_\rho)=(2,0)$, $(1,1)$ and $(0,2)$.
{In the $(2,0)$ and $(0,2)$ cases, the diquark spin should be 0, and
the} total baryon spin can be 
$J=3/2$, $5/2$ {as,}
\begin{eqnarray}
  \Lambda_c^*(J^+;\ell_\lambda\!=\!2)=
\left[
[\psi_{{0d}}(\vec{\lambda})\psi_{{0s}}(\vec{\rho}),
d^0]^{2},
\chi_c\right]^{J=2\pm 1/2} D^0c \ , \nonumber \\ \\
  \Lambda_c^*(J^+;\ell_\rho\!=\!2)=
\left[
[\psi_{{0s}}(\vec{\lambda})\psi_{{0d}}(\vec{\rho}),
d^0]^{2},
\chi_c\right]^{J=2\pm 1/2} D^0c \ , \nonumber \\
\end{eqnarray}
%
In the case with $(\ell_\lambda,\ell_\rho)=(1,1)$, the diquark spin
should be 1 {as}
\begin{eqnarray}
   \Lambda_c^*(J^+;\ell_\lambda=1,\ell_\rho=1)=
\left[
[\psi_{{0p}}(\vec{\lambda})\psi_{{0p}}(\vec{\rho}),
d^1]^{{j}},
\chi_c\right]^{{J}} D^0c \ . \nonumber \\
\end{eqnarray}
{The total angular momentum $\ell$
($\vec{\ell}=\vec{\ell}_\lambda+\vec{\ell}_\rho$) can be 0, 
1 and 2, and the {resulting} brown muck spin can be $j=(1)$, $(0, 1, 2)$, and $(1,
2, 3)$ giving 13 states.
{The heavy baryons are the HQS singlet only for $j=0$ and the HQS doublet for the others.}
}


{
We leave a comment on the difference between the wave function used in Ref.~\cite{Zhong:2007gp} and ours.
In Ref.~\cite{Zhong:2007gp}, the bases of the quark wave function are given by $^{2s+1}\ell_{J}$, namely
\begin{eqnarray}
 \left[\left[\ell_{\lambda}\ell_{\rho}]^{\ell}[[s_{1}s_{2}]s_{3}\right]^{s}\right]^{J},
\end{eqnarray}
while ours are given by
\begin{eqnarray}
 \left[\left[[\ell_{\lambda}\ell_{\rho}]^{\ell}[s_{1}s_{2}]^{s_{12}}\right]^{j}s_{3}\right]^{J}.
\end{eqnarray}
They are different in general except for the highest weight state of $\ell$ and $s$.
In the latter, the subcomponent $\left[[\ell_{\lambda}\ell_{\rho}]^{\ell}[s_{1}s_{2}]^{s_{12}}\right]^{j}$, which is assigned as the brown muck spin $j$, decouples from the heavy quark spin $s_{3}$ in the heavy quark limit.
Hence the latter basis is compatible with the heavy quark symmetry.
}

\section{Formulation}
\label{sec:formalism}
\subsection{Basic interaction of the pion}

{In the constituent quark model, the pion can couple to a single
quark through the Yukawa interaction, which is considered to contribute
dominantly to {one-}pion emission decays (Fig.~\ref{fig:decay}). In the
relativistic description, there are two independent couplings of
pseudo-scalar and {axial}-vector types,
\begin{equation}
 \bar{q}\gamma_5\vec{\tau}q\!\cdot\!\vec{\pi}, \ \ 
 \bar{q}\gamma_\mu\gamma_5\vec{\tau}q\!\cdot\! \del^\mu\vec{\pi}  \ .
\label{eq:2types}
\end{equation}
In the non-relativistic model, they correspond to the following two terms,}
\begin{equation}
 \vec{\sigma}\!\cdot\!(\vec{p}_i+\vec{p}_f)=\vec{\sigma}\!\cdot\! \vec{q}, \ \ 
 \vec{\sigma}\!\cdot\!(\vec{p}_i-\vec{p}_f) \ ,
\end{equation}
where $\vec{p}_i$ $(\vec{p}_f)$ is the momentum of the initial (final)
quarks and $\vec{q}$ is the pion momentum.
{We keep in mind that} these two couplings in Eq.~(\ref{eq:2types}) are
equivalent for the 
on-shell particles in the 
initial and final states, but not for the off-shell particles confined
within a finite size.  
{The present case is the latter, because the quarks are confined in the harmonic oscillator potential.} 
In this work, we employ the axial-{vector type} coupling,
\begin{equation}
 {\cal L}{_{\pi qq}}(x)= \frac{g_A^q}{{2f_\pi}}\bar{q}(x)\gamma_\mu\gamma_5 \vec{\tau}q(x)
\!\cdot\!\del^\mu \vec{\pi}(x),
\label{eq:L}
\end{equation}
in accordance with the low-energy chiral dynamics. 
{The non-relativistic limit in Eq.~(\ref{eq:L}) leads to {the combination of} the two terms in Eq.~(\ref{eq:2types}).}
In Eq.~(\ref{eq:L})}, $g_A^q$ is the axial coupling of the light quarks, for
which we use 
the value $g_A^q=1$~\cite{Weinberg:1990xm,Weinberg:1991gf}.  
{As we will see later, {importantly,} the {axial-}vector coupling can explain
surprisingly well the decay of $\Lambda_c^*(2595)$ through the 
time-derivative piece in Eq.~(\ref{eq:L}).  Contrary, the pseudoscalar 
coupling cannot reproduce it because it is proportional to the pion
momentum $q$ which almost vanishes.  This strongly supports the chiral
dynamics of the pion working with constituent light quarks.} 

\subsection{Matrix elements with the quark model wave functions}
{In this section, we formulate the one-pion emission decay of a charmed
baryon {within} the quark model.}
The relevant diagram is shown in Fig.~\ref{fig:decay}, where 
{one pion is emitted from a single {light} quark.}
We write state vector for the $Y_c$ baryon {($Y_c=\Lambda_c$ or
$\Sigma_c$)} with mass ${M_{Y_c}}$, 
{spin $J$ and momentum $P$ in the baryon rest frame 
{in the momentum representation}
as,}
\begin{multline}
 |Y_c{(P,J)}\ra =
  \sqrt{2M_{{Y_{{c}}}}}
\sum_{\{s,\ell \}}
\int \frac{d^3 p_\rho}{(2\pi)^3}\int \frac{d^3 p_\lambda}{(2\pi)^3}\\
\frac{1}{\sqrt{2m}}\frac{1}{\sqrt{2m}}\frac{1}{\sqrt{2M}}
\psi_{\ell_\rho}(\vec{p}_\rho)
\psi_{\ell_\lambda}(\vec{p}_\lambda)\\
|q_1(p_1,s_1)\ra |q_2(p_2,s_2)\ra|q_3(p_3,s_3)\ra.
\label{eq:B-ket}
\end{multline} 
{which is} a superposition of {quarks in the momentum space}
$|q_1(p_1,s_1)\ra$, $|q_2(p_2,s_2)\ra$,
and $|q_3(p_3,s_3)\ra$,  weighted by
the 
baryon wave functions $\psi{_{\rho}}(\vec{p}_\rho)$ and
$\psi{_{\lambda}}(\vec{p}_\lambda)$. {Here} the relative momenta 
{$\vec{p}_\rho$} and {$\vec{p}_\lambda$} are defined by
\begin{eqnarray}
  {\vec{p}}_\lambda&=&\dfrac{1}{2m+M}(M{\vec{p}}_1+M{\vec{p}}_2-2m{\vec{p}}_3)\\
  {\vec{p}}_\rho&=&\dfrac{1}{2}({\vec{p}}_1-{\vec{p}}_2) \ .
\end{eqnarray}
{and the total momentum of three quarks, which is the baryon
momentum, is given by
\begin{equation}
 {\vec{P}}={\vec{p}}_1+{\vec{p}}_2+{\vec{p}}_3 \ .
\end{equation}
}
 
\begin{figure}[hb]
 \begin{center}
  \includegraphics[width=0.6\linewidth]{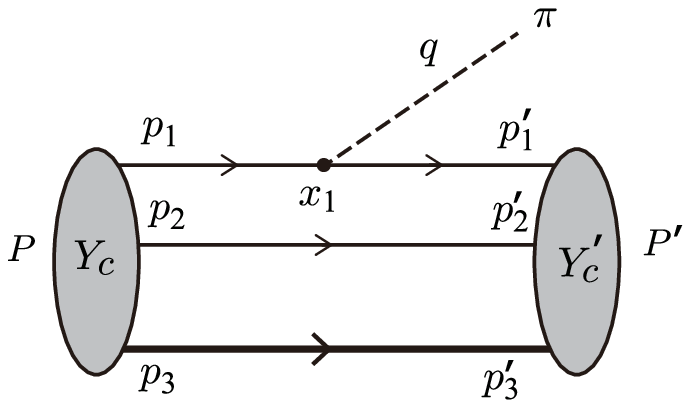}
 \end{center}
\caption{Decay amplitude of the charmed baryon $Y_c$ to
 $Y^{\prime}_c$ with one-pion emission.  
\label{fig:decay}}
\end{figure}
The factors of $1/\sqrt{2m}$ are for the normalizations of the
{confined} quark
states so that $\dis \int \frac{d^3 p_j}{(2\pi)^3}
|\psi(\vec{p}_j)|^2=1 $. 
The sum $\sum_{\{s,\ell\}}$ is taken over the spins
of the three quarks and 
their angular {momenta such} that the total angular momentum 
gives the spin $J$.

The decay amplitude for $Y_c \rightarrow Y'_c \pi$ is given by
\begin{equation}
 \int d^4 x_1 \la Y'_c{(P',J')} \pi{(q)} | i{\cal L}(x_1) | Y_c{(P,J)} \ra ,
\end{equation}
where only one light quark $|q_1\ra$ in the initial and final baryon state
participates in the transition as
\begin{multline}
\la q_1^{\prime}(p_1^{\prime},s_1^{\prime})\pi(q)|i{\cal L}{_{\pi qq}}(x_1)|q_1(p_1,s_1)\ra \simeq i\frac{g_A^q}{2\pi} e^{i(p_1'-p_1+q)\cdot x_1} \times
\\
\left\{i\omega_\pi\la \chi_{s_1^{\prime}}|(\vec{p}_1+\vec{p}^{\,\prime}_1)\!\cdot\!\vec{\sigma}|\chi_{s_1}\ra
-i2m\la \chi_{s^{\prime}_1}|(\vec{p}_1-\vec{p}^{\,\prime}_1)\!\cdot\!\vec{\sigma}|\chi_{s_1}\ra
\right\}, 
\label{eq:q->qpi}
\end{multline}
while the other light quark $|q_2\ra$ and the charm quark
$|q_3\ra$ are spectators and then their matrix elements
are just delta-functions of their three-momenta
\begin{eqnarray}
 \la
  q_j^{\prime}(p_j^{\prime},s_j^{\prime})|q_j(p_j,s_j)\ra
  &=&
2E_j(2\pi)^3\delta^{(3)}({\vec{p}}^{\,\prime}_j-\vec{p}_j)  
  \delta_{s_js_j^{\prime}} \nonumber \\
&=& 2E_j\int d^3 x_j e^{-i(\vec{p}^{\,\prime}_j-\vec{p}_j)\cdot \vec{x}_j}
\la \chi_{s_j^{\prime}}|\chi_{s_j}\ra \ , \nonumber \\
\end{eqnarray} 
where $j=2$ or $3$.  We have now ten $x$-integrals as
\begin{multline}
 \int dx_1^0 d^3 x_1 d^3 x_2 d^3 x_3 \\
e^{i(p_1'-p_1+q)\cdot x_1} 
e^{-i(\vec{p}^{\,\prime}_2-\vec{p}_2)\cdot \vec{x}_2}
e^{-i(\vec{p}^{\,\prime}_3-\vec{p}_3)\cdot \vec{x}_3} \ ,
\end{multline} 
and the first $x_1^0$-integral leads to the energy conservation $
 (2\pi)\delta(E_1-E_1'-\omega_\pi)
$ in $q_1 \rightarrow q_1'\pi$ process.  We rewrite the remaining
 $\vec{x}$-integrals  in terms of the Jacobi coordinates and we find
\begin{multline}
\int d^3 X d^3 \rho \, d^3 \lambda
e^{-i(\vec{P}^{\prime}-\vec{P})\cdot \vec{X}}
e^{-i(\vec{p}_\rho^{\,\prime}-\vec{p}_\rho)\cdot \vec{\rho}}
e^{-i(\vec{p}_\lambda^{\,\prime}-\vec{p}_\lambda)\cdot \vec{\lambda}}\\
{\times}
e^{-i\vec{q}\cdot(\vec{X}+\frac{M}{2m+M}\vec{\lambda}+\frac{1}{2}\vec{\rho})}.
\end{multline} 
The $\vec{X}$-integral leads to the total three-momentum conservation,
via $(2\pi)^3\delta^{(3)}(\vec{P}-\vec{P}'-\vec{q})$.
By eliminating the common delta-functions for the energy-momentum
conservation, we find the amplitude for $Y_c \rightarrow
Y'_c \pi$ decay as
\begin{widetext}
\begin{multline}
 -it_{Y_c \rightarrow Y'_c\pi}=\sum_{\{\Lambda,\Sigma\}} i\frac{g_A^q}{2f_\pi}\sqrt{2M_{Y_c}}\sqrt{2M_{Y'_c}}
\frac{1}{2m}
\int \frac{d^3p_\rho}{(2\pi)^3}
\int \frac{d^3p'_\rho}{(2\pi)^3}
\int \frac{d^3p_\lambda}{(2\pi)^3}
\int \frac{d^3p'_\lambda}{(2\pi)^3}
\int d^3 {\lambda}
\int d^3 {\rho}\\
\psi_{\ell_\rho'}^*(\vec{p}_\rho^{\,\prime})
e^{-i\vec{p}\,'_\rho\cdot\vec{\rho}}
\psi_{\ell_\rho}(\vec{p}_\rho)
e^{i\vec{p}_\rho\cdot\vec{\rho}}
\psi_{\ell_\lambda'}^*(\vec{p}_\lambda^{\,\prime})
e^{-i\vec{p}\,'_\lambda\cdot\vec{\lambda}}
\psi_{\ell_\lambda}(\vec{p}_\lambda)
e^{i\vec{p}_\lambda\cdot\vec{\lambda}}
e^{-i\vec{q}_\lambda\cdot\vec{\lambda}}
e^{-i\vec{q}_\rho\cdot\vec{\rho}}\\
\left\{i\omega_\pi\la \chi_{s^{\prime}_1}|(\vec{p}_\lambda^{\,\prime}+2\vec{p}^{\,\prime}_\rho)\!\cdot\!\vec{\sigma}|\chi_{s_1}\ra
+i\left(\omega_\pi\frac{M}{2m+M}-2m\right)\la \chi_{s^{\prime}_1}|
\vec{\sigma}\!\cdot\!\vec{q}\,|\chi_{s_1}\ra
\right\}
\la \chi_{s^{\prime}_2}|\chi_{s_2}\ra
\la \chi_{s^{\prime}_c}|\chi_{s_c}\ra \ ,
\label{eq:t}
\end{multline}
\end{widetext}
where the effective momentum transfer $\vec{q}_\lambda$ and
$\vec{q}_\rho$ appearing in the pion plain wave
$e^{-i\vec{q}\cdot\vec{x}_1}$ are defined by
\begin{equation}
 \vec{q}_\lambda = \frac{M}{2m+M}\vec{q}, \ \ 
 \vec{q}_\rho = \frac{1}{2}\vec{q} \ .
\end{equation}
The first term in Eq.~(\ref{eq:t}) involves the relative momenta
$\vec{p}_\rho^{\,\prime}$ and $\vec{p}_\lambda^{\,\prime}$ of the constituent quarks in the
final baryon, which can be replaced by the derivative of the wave
functions as  
\begin{eqnarray}
 \int \dfrac{d^3 p_\rho^{\,\prime}}{(2\pi)^3} \vec{p}_\rho^{\,\prime}
  \psi_{\ell_\rho'}^*(\vec{p}_\rho^{\,\prime}) e^{-i\vec{p}_\rho^{\,\prime}\cdot\vec{\rho}}
&=& i\vec{\nabla}_\rho
\int \dfrac{d^3 p_\rho^{\,\prime}}{(2\pi)^3} 
  \psi_{\ell_\rho'}^*(\vec{p}_\rho^{\,\prime}) e^{-i\vec{p}_\rho^{\,\prime}\cdot\vec{\rho}}
  \nonumber \\
&=& i\vec{\nabla}_\rho \psi_{\ell_\rho'}^*(\vec{\rho}\,),
\end{eqnarray}
and the same for $\vec{p}_\lambda^{\,\prime}$.   In the case of
$\Lambda_c(JM)^+\rightarrow \Sigma_c(J'M')^{++}\pi^-$,
{after performing the momentum integrals and by showing}
the flavor
(isospin) part explicitly, the decay amplitude is {given by}
\begin{widetext}
\begin{multline}
 -it_{\Lambda_c^+ \rightarrow
 \Sigma_c^{++}\pi^-}=
 -\frac{g_A^q}{2f_\pi}\sqrt{2M_{\Lambda_c}}\sqrt{2M_{\Sigma_c}} 
\frac{1}{2m}
\int d^3 \lambda \,
d^3 \rho \, e^{-i\vec{q}_\lambda\cdot\vec{\lambda}}
e^{-i\vec{q}_\rho\cdot\vec{\rho}} \left<D^1c\right|\tau^+_{(1)}\left|D^0c\right>
\\
\left<
 [[\psi_{\ell_\lambda}(\lambda)\psi_{\ell_\rho}(\rho),d]^{j'},\chi_c]^{J'}_{M'}\right| 
\left\{
\omega_\pi
(i\overleftarrow{\nabla}_\lambda+2i\overleftarrow{\nabla}_\rho)\!\cdot\!\vec{\sigma}_{(1)}
+
\left(\omega_\pi\frac{M}{2m+M}-2m\right)
\vec{\sigma}_{(1)}\!\cdot\!\vec{q} \right\} 
\left|[[\psi_{\ell_\lambda}(\lambda)\psi_{\ell_\rho}(\rho),d]^{j},\chi_c]^{J}_{M}
\right>,
\label{eq:t_2}
\end{multline}
\end{widetext}
where $\vec{\sigma}_{(1)}$ and $\tau^+_{(1)}$ matrices operate the
spin and isospin wave function{s} of the quark-1. 
For simplicity, the notation for the bra and ket states
\begin{multline}
 \left|[[\psi_{\ell_\lambda}(\lambda)\psi_{\ell_\rho}(\rho),d]^{j},\chi_c]^{J}_{M}
\right> \\
\equiv \sum_{\{\ell,s\}} \psi_{\ell_\lambda}(\lambda)\psi_{\ell_\rho} (\rho)
\l| \chi_{s_1} \>\l| \chi_{s_2} \>\l| \chi_{s_c} \>,
\end{multline}
are used in Eq.~(\ref{eq:t_2}). The derivatives
$\overleftarrow{\nabla}_\lambda$ and $\overleftarrow{\nabla}_\rho$
operate the final state wave functions.  
We also have to consider the case that the pion couples to 
{the another light quark $q_2(x_2)$.
Summing over the amplitudes of the two {cases} coherently, we obtain
the total decay amplitude.}

\subsection{Decay widths with the helicity amplitude}
{The decay width of $B_i\rightarrow B_f\pi$ is given by}
\begin{equation}
 \Gamma=\frac{1}{16\pi^2} \frac{q}{2M_i^2}\int d\Omega \sum_{f}
  \left|t_{{B_i\rightarrow B_f\pi}}\right|^2,
\label{eq:Gamma}
\end{equation}
where $q$ is the magnitude of the three-momentum of the final pion in
the center-of-mass frame, and the sum is {taken} over the possible
{quantum numbers}, in the present case, the spin state
(helicity) of the final baryon for a given initial baryon spin.
The matrix element depends
on the decay angle $\Omega$ (the angle between the quantization axis of the
initial baryon spin and {the} momentum vector $\vec{p}_f$ of the final baryon)
and on the helicity of $B_f$.  In this article, we employ the helicity
amplitude approach to calculate the decay width in Eq.~(\ref{eq:Gamma}).

In this approach, we expand the initial spin state
{$\l|B_i(J{,J_{z'}=}J)\>_{z'}$}, {which is} quantized 
along a fixed $\hat{e}_{z'}$ axis,
in the angular momentum
basis quantized along the direction of the momentum of the final baryon,
$\hat{e}_z={\vec{p}_f/|\vec{p}_f|}$, by
\begin{equation}
 \l|B_i(J,J)\>_{z'}=\sum_M \l|B_i(J,M)\>_z D^J_{MJ}(-\phi,\theta,\phi) \ ,
\end{equation}
where $D_{MJ}^J$ {are} the Wigner's $D$ functions~\cite{Morrison:1987}.
{If the {spin of the} final state $\<B_f(\vec{p}_f,h)\r|$
is quantized along $\hat{e}_z$, then the
helicity $h$ is {equal to} the third component of the final state spin,
\begin{equation}
 \<B_f(\vec{p}_f,h)\r| = {}_z\hspace{-2pt}\<B_f(\vec{p}_f,J',h)\r|,
\end{equation}
where $J'$ is the spin of the final baryon $B_f$.}

Hence the matrix element is written with its angular dependence as shown
explicitly
 \begin{multline}
{}_z\hspace{-2pt}\<B_f(\vec{p}_f,{J'},h)\pi(-\vec{p}_f)\r|\hat{t}
\l|B_i(J,J)\>_{z'} \\
=
{D^J_{MJ}(-\phi,\theta,\phi)
\hspace{1.5pt}{}_z\hspace{-2pt}\<B_f(\vec{p}_f,J',h)\pi(-\vec{p}_f) \r|
\hat{t}\l|B_i(J,h)\>_z
},
\label{eq:exp}
\end{multline}  
where only the diagonal element $M=h$ remains  after summing {over}
$\sum_M$, because {of} the helicity (spin $z$-component)
{conservation.} In Eq.~(\ref{eq:exp}) both of the initial and final
spins are quantized along $\hat{e}_z$ axis.

{Now, the helicity amplitude $A_h$ is} defined by
\begin{multline}
(2\pi)^4\delta^{(4)}(P_f-P_i)A_h\\
={}_z\hspace{-2pt}\<B_f(\vec{p}_f,{J'},h)\pi(-\vec{p}_f)\r|\hat{t}
\l|B_i(J,h)\>_z  \ .
\label{eq:Ah}
\end{multline}
The amplitude $A_h$ depends on $J$, $J'$ and $h$,
{but} does not depend on the decay angle,
because the spin quantization axis is chosen along the direction of the
momentum of the final baryon $\vec{p}_f$,
which is equal to the situation of the decay into the
$z$-direction.
The possible angular dependence of $\vec{p}_f$ is taken care of by the
$D$ function, and the angular-integral $d\Omega$ in Eq.~(\ref{eq:Gamma})
then can be performed exactly and {finally} we find 
\begin{equation}
 \Gamma=\frac{1}{4\pi} \frac{q}{2M_i^2} \frac{1}{2J+1} \sum_{h}
  \left|A_h\right|^2,
\label{eq:wid}
\end{equation}
where $q=|\vec{p}_f|$.  
{Here, the amplitude $A_{-h}$ with the opposite helicity has the same
form as $A_{h}$.}

%
%

\subsection{Parameters}
\label{sec:prmtr}
In the present Hamiltonian of the harmonic oscillator {in
Eq.~(\ref{eq:H})}, we have three 
model-parameters; $m$ the mass of the light quark, $M$ that of the heavy
quark, and $k$ the {spring constant}.  The masses of the
quarks {are set to be as,}
\begin{equation}
 m=0.35 \pm 0.05 \ {\rm (GeV)},
 M=1.5 \pm 0.1 \ {\rm (GeV)}.
\end{equation}
We {tune the value of $k$ so that} the level spacing of the 
$\lambda$-mode excitation {as}
$\omega_\lambda \sim 0.35\pm 0.05$~GeV
and the root-mean-square
radius of the charmed baryon {as} $\sqrt{\left<R^2\right>}\sim$ 0.45--0.55~fm
{which is defined as the average of the distance of each quark from the
center-of-mass as,
\begin{eqnarray}
R^2 &\equiv &\dfrac{1}{3} \sum_{i=1}^3\left(\vec{r}_i-\vec{X}\right)^2
 \nonumber \\
 &= &\dfrac{1}{3}\left(\dfrac{2(2m^2+M^2)}{(2m+M)^2}\lambda^2
+\dfrac{1}{2}\rho^2
\right), 
\end{eqnarray}}
We summarize the model parameters used in the present calculation in Table
\ref{tab:prmtr}.  
{Depending on these input parameters,}
the range parameters of the Gaussian
wave functions {vary within the range of}
\begin{equation}
 a_\lambda=0.36 \text{--}0.44~{\rm (GeV)}, \ \ 
 a_\rho = 0.26 \text{--}0.32~{\rm (GeV)},
\end{equation}
{which is the source of the uncertainty in our theory predictions.}

\begin{table}[htb]
\caption{{Range of the m}odel parameters {of $\{m,M,k\}$
 (inputs) and the properties of resulting harmonic oscillator functions (outputs).}
\label{tab:prmtr}}
\begin{center}
\begin{tabular}{ccc} \hline\hline
 & light quark mass $m$ & 0.3--0.4~(GeV) \\
inputs & heavy quark mass $M$ & 1.4--{1.6}~(GeV) \\
 & H.O. potential $k$ & 0.02--{0.038}~(GeV$^3$) \\\hline
 & H.O. energy $\omega_\lambda$ & 0.3--0.4~(GeV) \\
outputs & H.O. energy $\omega_\rho$ & 0.42--0.58~(GeV) \\
 & gauss range $a_\lambda$ & 0.36--0.44~(GeV) \\
 & gauss range $a_\rho$ & 0.26--0.32~(GeV) \\
 & $\sqrt{\left<\lambda^2\right>}$ & 0.55--0.67~(fm) \\
 & $\sqrt{\left<\rho^2\right>}$ & 0.76--0.93~(fm) \\
 & $\sqrt{\left<R^2\right>}$ & 0.45--0.55 (fm) \\
\end{tabular}
\end{center}
\end{table}

 \section{Numerical Results}
\label{sec:results}
\subsection{Decays of the ground state $\Sigma_c(1/2^+)$  and
$\Sigma_c^*(3/2^+) \rightarrow \Lambda_c(1/2^+)\pi$}
\label{sec:Sigma}
\begin{table*}[htbp]
\caption{Calculated decay widths of $\Sigma_c(2455)^{++}$ and
 $\Sigma_c^*(2520)^{++}$ into the $\Lambda_c(2286)^+\pi^+$ {pair}. 
{$q$ is the momentum of the final particle in center-of-mass frame.}}
\label{tab:gs}
\begin{center}
\begin{tabular}{c|c|c||c} \hline
$B_i$ $J^P$ & $\Gamma_{\rm exp}$ & $q$ & $\Gamma_{\rm th} (\Sigma_c(J^+)^{++} \rightarrow \Lambda_c(2286)^{+}\pi^+)$ \\
(MeV) & (MeV) & (MeV/c) & (MeV) \\\hline\hline
$\Sigma_c(2455)^{++}$ $1/2^+$ & {1.89} & 89 & 4.27--4.33 \\
$(2453.98)$ & \ &  &  \\
 &  &  &  \\
$\Sigma_c^*(2520)^{++}$ $3/2^+$ & {14.78} & {177} & {30.3--31.6} \\
$(2517.9)$ &  &  &  \\
\end{tabular}
\end{center}
\end{table*}

The $\Sigma_c(2455)$ baryon is an orbital ground state baryon having
$J^P=1/2^+$.  The mass of the $\Sigma_c(2455)^{++}$ is {$ 2453.97 \pm
0.14$~MeV} and its full width 
is
$1.89^{+0.09}_{-0.18}$~(MeV)~\cite{Agashe:2014kda}. 
The $\Sigma_c(2455)\rightarrow\Lambda_c(2286)\pi$ {decay} channel is the only
possible strong decay and its branching ratio is $\sim 100\%$.  
{The $\Sigma_c^*(2520)$ baryon has $J^P=3/2^+$ and is
expected to form a {HQS} doublet with $\Sigma_c(2455)$.  The mass of
the $\Sigma_c^*(2520)^{++}$ is $2518.41^{+0.21}_{-0.19}$~(MeV) and its
width is $14.78^{+0.30}_{-0.40}$~(MeV)~\cite{Agashe:2014kda}. Again 
the $\Lambda_c(2286)\pi$ decay channel is the only
possible channel in the strong decay and its branching ratio is $\sim 100\%$.  
}
{Because both $\Sigma_c(2455)$ and $\Sigma_{c}^{\ast}(2520)$  baryons
are the spin and isospin 
flip states of the ground state $\Lambda_c(2286)$, their decay rates
reflect mainly the spin-isospin structure and is rather insensitive to
the spatial structure.  Therefore, we can use these processes to check
the validity of the present quark model calculations.}

The helicity amplitude for the
$\Sigma_c(1/2^+)\rightarrow\Lambda_c(1/2^+)\pi$ decay is {given by},
\begin{equation}
 A_h=A_h^{\nabla\cdot\sigma}+A_h^{q\cdot,
  \sigma},
\end{equation}
where $A_h^{\nabla\cdot\sigma}$ and $A_h^{q\cdot\sigma}$
correspond to
 the $(\vec{\nabla}_\lambda+2\vec{\nabla}_\rho)\cdot\vec{\sigma}$ term 
and
 the $\vec{q}\cdot\vec{\sigma}$ term
in Eq.~(\ref{eq:t_2}), respectively.
They are {given} by
\begin{eqnarray}
-iA_{1/2}^{\nabla\cdot\sigma}=G\frac{\omega_\pi}{m} \left(-\frac{1}{\sqrt{3}}\right)
\left(\frac{1}{2}q_\lambda+q_\rho\right)F(q),
\label{eq:Ahn(Sigma)}
\end{eqnarray}
and
\begin{eqnarray}
 -iA_{1/2}^{q\cdot\sigma}=-G\frac{q}{m}
\left(-\frac{1}{\sqrt{3}}\right)
\left(\frac{M}{2m+M}\omega_\pi-2m\right)F(q), \nonumber \\
\label{eq:Ahs(Sigma)}
\end{eqnarray}
where $q_{\lambda(\rho)}\equiv|\vec{q}_{\lambda(\rho)}|$ and
$G$ denotes the coupling constant and the normalizations as,
\begin{equation}
 G = \frac{g_A^q}{2f_\pi}\sqrt{2M_{\Lambda_{{c}}}}\sqrt{2M_{\Sigma_c}} \ .
\end{equation}
The function $F(q)$ denotes the Gaussian form factor as
\begin{equation}
 F(q)=e^{-{q_\lambda^2}/{4a_\lambda^2}}e^{-{q_\rho^2}/{4a_\rho^2}}\ ,
\end{equation}
which {is the Fourier transform of ground to ground transition
amplitude.}
The factors of $a_\lambda$ and $a_\rho$ correspond to the inverse of the
range of the 
Gaussian wave functions for $\lambda$- and $\rho$-motions, respectively,
and their definitions are given in Appendix.
Similarly, the helicity amplitude for the
$\Sigma_c^*(3/2^+)\rightarrow\Lambda_c(1/2^+)\pi$ decay is given by the
same expressions as Eqs.~(\ref{eq:Ahn(Sigma)}) and (\ref{eq:Ahs(Sigma)})
but the factor $-1/\sqrt{3}$ is replaced by $\sqrt{2/3}$ in both equations.

In Table~\ref{tab:gs}, we show the numerical results for the
$\Sigma_c(2455)(1/2^+)^{++}\rightarrow\Lambda_c^{+}\pi^{+}$ decay.
The calculated decay width is almost {twice} {as large as} the
experimental value.  We also show the {results of}
$\Sigma_c^*(2520)(3/2^+)$ decay in the same table.
The calculated decay width of
$\Sigma_c^*(3/2^+)$ is again {twice} {as large as}
the experimental value.

As shown in the table, the uncertainty from the ambiguities of the quark
model parameters $(m, M, k)$ is small, which means the decay width of
the ground state to the ground state does not depend on the detail of
the wave functions, as anticipated.   Therefore the discrepancy might come from the
axial-coupling constant $g_A^q$ for the $\pi qq$ interaction.

In the present calculation, we employ $g_A^q=1$ for the $q\pi\pi$
coupling, but it is also known that this value does not reproduce the
axial-coupling constant of the nucleon $g_A^{N}=1.25$ but leads to
$g_A^{N}=5/3$ instead.  To reproduce the axial-coupling constant of the
nucleon $g_A^{N}$, one needs {a suppression} factor of {about} 3/4
for $g_A^q$, 
which {reduces the decay width by a}
factor $(3/4)^2\sim 0.56$,
{the result of which is consistent with the experimental data.}
{This is expected because the pion emission decays essentially
measure the axial couplings for relevant baryons (transitions).   
Our input here is the axial coupling of the constituent quarks which can
take in principle any value when chiral symmetry is spontaneously
broken.   
Here we have shown that it is ~3/4 empirically from the phenomena of the
ground state baryons not only for the nucleon but also for charmed
baryons, which is not {far from} the discussion of
Weinberg~\cite{Weinberg:1990xm}. 
{The suppression of $g_A$ has been considered to be originated from
the mixing of $p$-waves due to relativistic corrections or pion
clouds~\cite{Chodos:1974pp}.}  
This, however, may vary for different
baryon excitations.  
Keeping this in mind, in the following calculations for decays of the
excited states, we keep using the value $g_A^q = 1$.     
}

\subsection{$\Lambda_c^*{(2595)(1/2^-)}\rightarrow\Sigma_c(2455)(1/2^+)\pi$}

The $\Lambda_c^*(2595)^+$ baryon is the first excited charmed baryons with
$I=0$ and is expected to have $J^P=1/2^-$.  The total decay width is
$\Gamma_{\rm exp}=2.6\pm 0.6$~MeV, where the $\Lambda_c^+\pi\pi$ channel is {the only
strong decay. The $\Lambda_c^+\pi\pi$ seems to be dominated by
$\Sigma_c(2455)\pi$ and its branching ratio
$\Gamma(\Sigma_c\pi)/\Gamma(\rm total)$ is quoted as
$BR(\Sigma_c^{++}\pi^-)= BR(\Sigma_c^{0}\pi^+)=24 \pm 7$ \%~\cite{Agashe:2014kda}.}
{The direct three-body decay width is $18\pm 10 \%$ which we do not
calculate in this article.}

Employing the quark model, we have three possibilities to describe the
excited $\Lambda_c^*$ baryon having $J^P=1/2^-$ as discussed in the
previous section.  One is the $\lambda$-mode excitation having $j^P=1^-$, and
the other two are the $\rho$-mode excitations having $j^P=0^-$ and $j^P=1^-$.

The helicity amplitude for the $\pi^-$ emission decay of
$\Lambda_c^*(1/2^-;\lambda)^+ \rightarrow \Sigma_c(1/2^+)^{++}\pi^-$ is
found again as the sum 
\begin{equation}
 A_h(1/2^-;\lambda)=A_h^{\nabla\cdot\sigma}(1/2^-;\lambda)+A_h^{q\cdot\sigma}(1/2^-;\lambda),
\end{equation}
where
\begin{multline}
 -iA_{1/2}^{\nabla\cdot\sigma}(1/2^-;\lambda)\\
=iG \dfrac{\omega_\pi}{m} \
\left\{
c_0 a_\lambda + c_2 \left(\dfrac{1}{2}q_\lambda+q_\rho\right)\dfrac{q_\lambda}{a_\lambda}
\right\}F(q),
\label{eq:Ahn(l)}
\end{multline}
and
\begin{multline}
-iA_{1/2}^{q\cdot\sigma}(1/2^-;\lambda)\\
=-iG\dfrac{q}{m}\left(\dfrac{M}{2m+M}\omega_\pi-2m\right)
c_2\dfrac{q_\lambda}{a_\lambda} F(q),
\label{eq:Ahs(l)}
\end{multline}
where 
\begin{equation}
c_0=-\frac{1}{\sqrt{2}}, \ \ c_2=\frac{1}{3\sqrt{2}},
\end{equation}
{which are determined by the Clebsch-Gordan coefficients.}
We summarize the general expressions in Appendix.
We can see that, the $A^{\nabla\cdot \sigma}$ starts from 
${\cal O}(q^0)$ {reflecting properly the nature of possible $s$-wave
decay}, while
$A^{q\cdot\sigma}$ is of order ${\cal O}(q^2)$. 
We will see that
the former gives a considerable contribution {to} the $\Lambda_c^*(2595)$
decay width.

As for the $\rho$-mode with $j=1$, we find a similar form for the
$\Lambda_c^*(1/2^-,\rho_{j=1})^+\rightarrow\Sigma(1/2^+)^{++}\pi^-$ decay as
\begin{multline}
 -iA_{1/2}^{\nabla\cdot\sigma}(1/2^-;\rho_{j=1})\\
=iG \dfrac{\omega_\pi}{m} \
\left\{
c_0 a_\rho + c_2 \left(\dfrac{1}{2}q_\lambda+q_\rho\right)\dfrac{q_\rho}{a_\rho}
\right\}F(q),
\label{eq:Ahn(r)}
\end{multline}
and
\begin{multline}
-iA_{1/2}^{q\cdot\sigma}(1/2^-;\rho_{j=1})\\
=-iG\dfrac{q}{m}\left(\dfrac{M}{2m+M}\omega_\pi-2m\right)
c_2\dfrac{q_\rho}{a_\rho} F(q),
\label{eq:Ahs(r)}
\end{multline}
where 
\begin{equation}
c_0=2, \ \ c_2=-\frac{1}{3}. 
\end{equation}

In contrast to {the above two cases}, the situation is quite
different for the decay 
of $\Lambda_c^*(1/2^-,\rho_{j=0})$ having the brown muck spin $j=0$.  The
amplitudes are exactly zero as,
\begin{gather}
 A_{1/2}^{\nabla\cdot\sigma}(1/2^-;\rho_{j=0})=0, \\
 A_{1/2}^{q\cdot\sigma}(1/2^-;\rho_{j=0})=0,
\end{gather} 
for the decay into $\Sigma_c(1/2^+)$ baryon.  This is due to {the spin
conservation
of the brown muck;} the
spin-parity $j^P=0^-$ state cannot decay into $j^P=1^+$ with the pion
$0^-$ {for any combination of relative angular momentum}. 
\label{sec:brown}
Generally, as we will see more examples, such requirements lead to
selection rules due to the consistency between the decays of baryons and
decays of brown muck, or the diquark in the quark model because the pion
couples only to the light quarks. 
{Such observations can be done best by using the baryon wave
functions as inspired by the heavy quark symmetry.}  

\begin{figure}[thb]
 \includegraphics[width=0.6\linewidth]{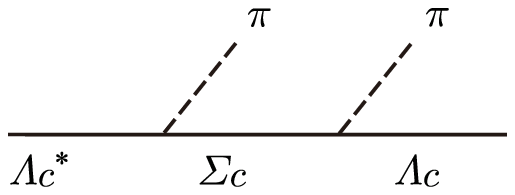}
\caption{Feynman diagram of the sequential decay of $\Lambda_c^*
 \rightarrow \Sigma_c\pi$ followed by
 $\Sigma_c\rightarrow\Lambda_c\pi$ supposed in
 Eq.~(\ref{eq:Con}). 
\label{fig:dia}}
\end{figure}

To {estimate} the decay width of the $\Lambda_c^*(2595)$
baryon, we should take the finite width of the finial $\Sigma_c$ baryon
into account,
because the $\Sigma_c\pi$ threshold is very close to the
$\Lambda_c^*(2595)$ mass.  
Indeed, the
$\Sigma_c^{++}\pi^-$ and $\Sigma_c^{0}\pi^+$
channels barely 
close at the $\Lambda_c^*(2595)$ mass while {the} $\Sigma_c^+\pi^0$
{channel} opens,
which means the isospin breaking is large contrary to the
{assumption made in PDG~\cite{Agashe:2014kda}.} 
To this end, we convolute the decay width of
$\Lambda_c^*(2595)$ by the finite width of $\Sigma_c$ as
\begin{eqnarray}
 \tilde{\Gamma}_{\Lambda^*_{{c}}}=
  \frac{1}{N}
\int d\tilde{M}_{\Sigma_{{c}}}\,
{\rm
Im}\frac{\Gamma_{\Lambda^*_{{c}}}(\tilde{M}_{\Sigma_{{c}}})}{\tilde{M}_{\Sigma_{{c}}}-M_{\Sigma_{{c}}}+i\Gamma_{\Sigma_{{c}}}(\tilde{M}_{\Sigma_{{c}}})/2}, \nonumber \\
\label{eq:Con}
\end{eqnarray}
where {$\Gamma_{\Lambda^*}(\tilde{M}_\Sigma)$ is the calculated
decay width of $\Lambda_c^*$ given in Eq.~(\ref{eq:wid}) which depends
on the mass $\tilde{M}_\Sigma$ of the final $\Sigma_c$
baryon. The normalization 
factor $N$ is defined by,}
\begin{equation}
N=
\int d\tilde{M}_{\Sigma_{{c}}}\,
{\rm Im}
\frac{1}{\tilde{M}_{\Sigma_{{c}}}-M_{\Sigma_{{c}}}+i\Gamma_{\Sigma_{{c}}}(\tilde{M}_{\Sigma_{{c}}})/2}\ .
\end{equation}
We take into account the phase space factor for the $\Sigma_c$ decay width in
the convolution integral as,
\begin{multline}
 \Gamma_{\Sigma}(\tilde{M}_{\Sigma_{{c}}})=
\Gamma_{\Sigma_{{c}}} \frac{M_{\Sigma_{{c}}}}{\tilde{M}_{\Sigma_{{c}}}}
\left(
\frac{\lambda^{1/2}(\tilde{M}_{\Sigma_c}^2,M_{\Lambda_c}^2,m_\pi^2)}
{\lambda^{1/2}(M^2_{\Sigma_{{c}}},M_{\Lambda_{{c}}}^2,m_\pi^2)}
\right)^3 \\
\times\theta(\tilde{M}_{\Sigma_{{c}}}-M_{\Lambda_{{c}}}-m_\pi),
\end{multline}
{where $M_{\Lambda_{{c}}}$ is the mass of the ground state
$\Lambda_c(2286)$, and $\Gamma_{\Sigma_{{c}}}$ is the decay width of $\Sigma_c$
given by}
$\Gamma_{\Sigma_{{c}}}=1.89$~(MeV) for $\Sigma_c^{++}$, 
$\Gamma_{\Sigma_{{c}}}=1.83$~(MeV) for $\Sigma_c^{0}$.  {Because only the
upper limit is determined for $\Sigma_c^{+}$,} we
calculate the ratio of $\Gamma(\Sigma_c^{++})/\Gamma(\Sigma_c^+)$ by
employing our formalism discussed in Sec.~\ref{sec:Sigma}, and then estimate it
as $\Gamma_{\Sigma_{{c}}}=2.1$~(MeV) for $\Sigma_c^+$.
{The convolution} corresponds to the consideration of the
sequential decay of the $\Lambda_c^*\rightarrow \Sigma_c\pi$
followed by $\Sigma_c\rightarrow\Lambda_c\pi$ as depicted in
Fig.~\ref{fig:dia}. 
{The double $\pi^0$ emission decay of
$\Lambda_c^*(2595)^+\rightarrow\Lambda_c(2286)\pi^0\pi^0$ can be
approximated by the convoluted single $\pi^0$ decay of
$\Lambda_c^*(2595)^+\rightarrow\Sigma_c(2455)^{+}\pi^0$
(including a symmetry factor for the two identical particles), because 
of the dominant contribution of the on-shell $\Sigma_c$~\cite{Cho:1994vg}.
Similarly, the 
charged pion decay $\Lambda_c\pi^+\pi^-$ is approximated by the sum
of the $\Sigma_c^{++}\pi^-$ and $\Sigma_c^0\pi^+$ decays.}

\begin{figure}[htb]
 \includegraphics[width=\linewidth]{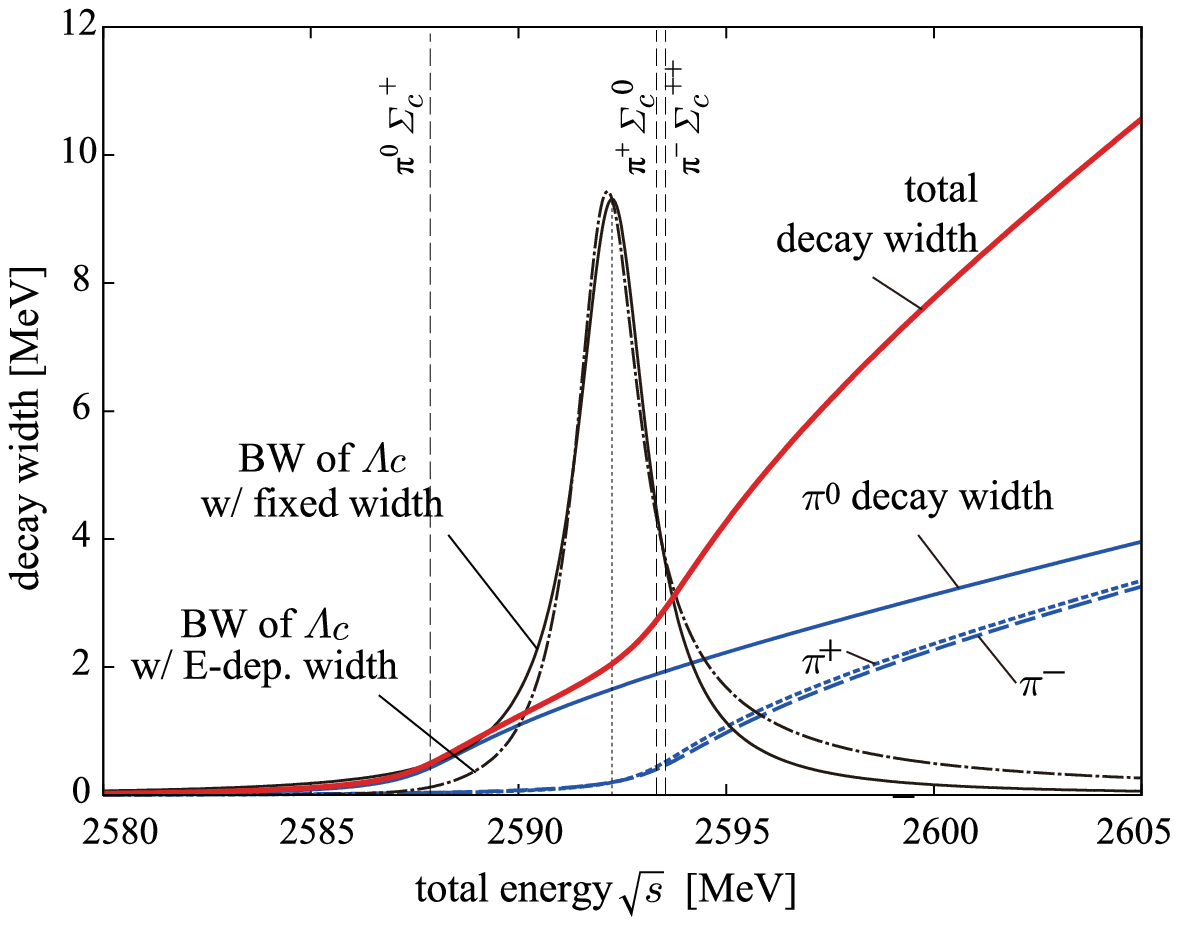}
\caption{(color online) Convoluted decay width of
 $\Lambda_c^*(2595;\lambda\text{-mode})\rightarrow\Sigma_c(2455)\pi$ as
 functions of total energy ($=$ the mass of the $\Lambda_c^*$).  
{The thin (blue) lines denote the $\pi^-$, $\pi^0$, and $\pi^+$
 emission decay widths as indicated in the figure.}
 The thick (red) solid line denotes {the sum of three charge states.}
The
 resulting Breit-Winger spectral functions of the $\Lambda_c^*$ are also
 shown {in arbitrary unit.}
\label{fig:amp}}
\end{figure}

\begin{table*}[hbt]
\caption{{Calculated decay width of the
 $\Lambda_c^*(2595)\rightarrow \Sigma_c(2455)\pi$.  The charge decay
 {channels} are indicated in the table, where $[\Sigma_c\pi]^+$ denotes the
 isospin summed width.
The quantum numbers of the $\lambda$- and $\rho$-modes are indicated by
$(n_\lambda,\ell_\lambda)$ and $(n_\rho,\ell_\rho)$, and
 $J_{\Lambda^{\ast}_{{c}}}(j)^P$ stands for the assigned spin and parity for
 $\Lambda_c^*$ with the brown muck spin $j$.
The masses of the $\Lambda_c^*$, $\Sigma_c$, and $\pi$ are also shown in the
 table.  The symbol $\dagger$ indicates the closed channels for on-shell
 $\Sigma_c\pi$. 
\label{tab:2595}}}
\begin{center}
\begin{tabular}{cccccccc} 
\multicolumn{7}{l}{$\Lambda_c^*(2595)^+$ decay width ($M_{\Lambda^*}=2592.25$~(MeV))} &  \\\hline\hline
\multicolumn{3}{r}{decay channel} & full & $[\Sigma_c\pi]^+$ & $\Sigma_c^{++}\pi^-$ & $\Sigma_c^{0}\pi^+$ & $\Sigma_c^{+}\pi^0$  \\
\multicolumn{3}{r}{Experimental value $\Gamma_{\rm exp}$ (MeV)~\cite{Agashe:2014kda}} & $2.6 \pm 0.6$ & - & 0.624 (24\%) & 0.624 (24\%) & - \\
\multicolumn{3}{r}{momentum of final particle $q$ (MeV/c)} & - & - & $\dagger$ & $\dagger$ & 29 \\\hline\hline
this work & $(n_\lambda,\ell_\lambda)$, $(n_\rho,\ell_\rho)$ & $J_\Lambda(j)^P$ &  &  &  &  &  \\\cline{2-8}
$\Gamma$ & $(0,1)$, $(0,0)$ & $1/2(1)^-$ &  & 1.5--2.9 & 0.13--0.25 & 0.15--0.28 & 1.2--2.4 \\\cline{2-8}
(MeV) & $(0,0)$, $(0,1)$ & $1/2(0)^-$ &  & 0 & 0 & 0 & 0 \\
 &  & $1/2(1)^-$ &  & 6.5--11.9 & 0.57--1.04 & 0.63--1.15 & 5.3--9.7 \\\hline\hline
 &  &  & \multicolumn{2}{c}{$M_{\Sigma}$ (MeV)} & 2453.97 & 2453.75 & 2452.9 \\
 & \multicolumn{2}{r}{input parameters employed} & \multicolumn{2}{c}{$\Gamma_{\Sigma}$ (MeV)} & 1.89 & 1.83 & 2.1 \\
 & \multicolumn{2}{r}{in the convolution Eq.~(\ref{eq:Con})} & \multicolumn{2}{c}{$m_\pi$ (MeV)} & 139.57 & 139.57 & 134.98 \\
\end{tabular}
\end{center}

\end{table*}

In Fig.~\ref{fig:amp}, we show the calculated result for the decay width
of the $\Lambda_c^*(2595)$ baryon in the case of the $\lambda$-mode
 as functions of the mass of the $\Lambda_c^*$ (the total energy $\sqrt{s}$).
We find that the $\pi^{ \pm}$ decay width remains finite even at
$\sqrt{s}=M_{\Lambda_c^*}$ which is below the $\pi^{ \pm}$ threshold, owing
to the finite width of the $\Sigma_c$ baryon. 
We can also see that the $\pi^0$ threshold is located at 5~MeV below
$\sqrt{s}=M_{\Lambda_c^*}$ and then the $\pi^0$ decay width is much
larger than that of $\pi^{{\pm}}$, meaning a large isospin breaking.
{We also show the resulting Breit-Wigner form in Fig.~\ref{fig:amp}
with the fixed width at $\sqrt{s}=M_{\Lambda^*_{{c}}}=2592.25$~(MeV) and with
the energy-dependent width.  In the present case, both of the BW functions
resemble because of the resulting small width.  However, the
energy-dependence {of the width} is large, so we have to be careful
when estimating the 
BW width for $\Lambda_c^*(2595)$.}

In Table \ref{tab:2595} we show the calculated decay
widths of {$\Lambda_c^*(2595)^+\rightarrow\Sigma_c(2455)^{++}\pi^-$,
$\Sigma_c(2455)^0\pi^+$, and $\Sigma_c(2455)^+\pi^0$ together with the
sum of these three channels}
{evaluated at $\sqrt{s}=M_{\Lambda^*_{c}}=2592.25$~(MeV)}. 
These numbers have uncertainty reflecting that of model parameters of
$(m, M, k)$ as
discussed in Sec.~\ref{sec:prmtr}.  The uncertainty of the model
parameters leads to almost {factor-two} difference in the decay widths.  
In spite of {this uncertainty} {including the one coming from
$g_A^q$},  
{using the axial-vector coupling works well}
to reproduce the relatively large
decay width of $\Lambda_c^*(2595)$ located at almost $\Sigma_c\pi$
threshold.
{This is due to the time derivative term with the strength
determined the mass of the pion.
Thus the decay of $\Lambda^*_c(2595)$ provides a good example to show
that the chiral theory works up to the order ${\cal O}(m_\pi)$.}
{As discussed in the previous section, we find that, 
by employing
the pseudo-scalar coupling ($\gamma_5$) 
for the pion, we obtain less than 1~(keV) for the $\Lambda_c^*(2595)$
decay due to the small pion momentum $q$.}

We also find that the assignment of the $\rho$-mode configuration with
$j^P=1^-$ to the $\Lambda_c^*(2595)$ leads to  almost 2.5 -- 5 times larger width than
the experimental value {for the total width}. {They are
significantly large}
even if we consider the uncertainty of the pion coupling, because 
the experimental total width contains not only the $\Sigma_c\pi$ decay channel
but also the non-resonant three-body decay of $\Lambda_c\pi\pi$
{which we do not consider in this paper.}

In addition, the
$\rho$-mode configuration with $j^P=0^-$ cannot decay into
$\Sigma_c\pi$.  Therefore we can conclude that, 
{by the detailed study of decay width},
it is
likely that
$\Lambda_c^*(2595)$ baryon is dominated by the $\lambda$-mode
configuration as expected.
We might add a comment that other assignments of the $J^P=3/2^-$ or higher spin
configurations for $\Lambda_c^*(2595)$ cannot reproduce the {large} experimental
value for the decay width {due to $d$-wave {(or higher partial wave)} nature.}

\subsection{$\Lambda_c^*(2625)(3/2^-)\rightarrow
  \Sigma_c(2455)(1/2^+)\pi$}

\begin{table}[bht]
\caption{{Calculated decay widths of the
 $\Lambda_c^*(2625)\rightarrow\Sigma_c(2455)^{++}\pi^-$. 
The quantum numbers of the $\lambda$- and $\rho$-mode are indicated by
$(n_\lambda,\ell_\lambda)$ and $(n_\rho,\ell_\rho)$, and
 $J_{\Lambda^{\ast}_{{c}}}(j)^P$ stands for the assigned spin and parity for $\Lambda_c^*$
with the brown muck spin $j$.  The masses of the
 $\Sigma_c^{++}$ and $\pi^-$ are $M_{\Sigma^{++}_c}=2453.97$~(MeV) and
 $m_{\pi^-}=139.57$~(MeV). }
\label{tab:2625}}
\begin{center}
\begin{tabular}{ccccc} 
\multicolumn{5}{l}{$\Lambda_c^*(2625)^+$ decay width  ($M_{\Lambda^*}=2628.11$~(MeV))} \\\hline\hline
\multicolumn{3}{r}{decay channel} & full & $\Sigma_c^{++}\pi^-$ \\
\multicolumn{3}{r}{Experimental value $\Gamma_{\rm exp}$ (MeV)~\cite{Agashe:2014kda}} & $< 0.97$ & $<0.05 (<5\%)$ \\
\multicolumn{3}{r}{momentum of final particle $q$ (MeV/c)} & - & 101 \\\hline\hline
this work & $(n_\lambda,\ell_\lambda)$, $(n_\rho,\ell_\rho)$ & $J_\Lambda(j)^P$ &  &  \\\cline{2-5}
$\Gamma$ & $(0,1)$, $(0,0)$ & $1/2(1)^-$ &  & 5.4--10.7 \\
(MeV) &  & $3/2(1)^-$ &  & 0.024--0.039 \\\cline{2-5}
 & $(0,0)$, $(0,1)$ & $1/2(0)^-$ &  & 0 \\\cline{3-5}
 &  & $1/2(1)^-$ &  & 24.0--45.1 \\
 &  & $3/2(1)^-$ &  & 0.013--0.019 \\\cline{3-5}
 &  & $3/2(2)^-$ &  & 0.023--0.034 \\
 &  & $5/2(2)^-$ &  & 0.010--0.015 \\
\end{tabular}
\end{center}

\end{table}

The $\Lambda_c^*(2625)^+$ baryon is very narrow resonant state and is
expected to have $J^P=3/2^-$.  
{In PDG, only the upper limit of {the decay width} is given as}
$\Gamma_{\rm exp} < 0.97$~MeV~\cite{Agashe:2014kda}.  The
$\Lambda_c^+\pi\pi$ and its submode $\Sigma_c\pi$ are the only strong
decay {channel}.  The branching ratio
$BR(\Sigma_c^{++}\pi^-)/BR(\Lambda_c^+\pi^+\pi^-)$ is less than 5\%, and
therefore the partial decay width for 
$\Gamma_{\rm exp}(\Lambda_c^*(2625)^+\rightarrow
\Sigma_c^{++}\pi^-)$ {is less than}  0.05~MeV.

As discussed in the previous section, the $\Lambda_c^*(2625)$ baryon is
{assigned to be} the
low-lying orbital excitation state with $\ell_\lambda=1$ with spin-0 light
diquark.
The helicity amplitude for the
$\Lambda_c^{{\ast}}(3/2^-;\lambda)^+\rightarrow\Sigma_c^{++}\pi^-$ is {then}
{given by the same expressions as Eqs.~(\ref{eq:Ahn(l)}) and
(\ref{eq:Ahs(l)}) but with the different coefficients as
\begin{equation}
 c_0=0, \ \ c_2=-\frac{1}{3} \ .
\end{equation}
}
In contrast to the {case of} $\Lambda_c^*(2595)$, the coefficient
{$c_0$} of the $q^0$ term is zero then
the helicity amplitudes
$A_h^{\nabla\cdot\sigma}$ and $ A_h^{q\cdot\sigma}$ are of
order of ${\cal O}(q^2)$ as expected {for} the $3/2^-\rightarrow
1/2^++0^-$ decay.

We have two more possible quark
configurations for the $\Lambda_c^*$ excitations with $J^P=3/2^-$,
which are the $\rho$-mode excitation{s} with $j=1$ and $j=2$.
The helicity amplitudes for these configurations are found {to be}
{again the same as Eqs.~(\ref{eq:Ahn(r)}) and
(\ref{eq:Ahs(r)}) but with different coefficients as
\begin{equation}
 c_0 = 0, \ \ c_2=-\frac{1}{3\sqrt{2}},
\end{equation}
for $\Lambda_c^*(3/2^-,\rho_{j=1})\rightarrow \Sigma_c(1/2^+)\pi$ decay,
and
\begin{equation}
 c_0 = 0, \ \ c_2=\frac{1}{\sqrt{10}},
\end{equation}
for $\Lambda_c^*(3/2^-,\rho_{j=2})\rightarrow \Sigma_c(1/2^+)\pi$ decay.}

In Table~\ref{tab:2625} we show the numerical results for the
$\Lambda_c^*(2625)^+\rightarrow \Sigma_c(2455)^{++}\pi^-$ decay.
In the $\Lambda_c^*(2625)$ case, we do not convolute {over the finite
width of $\Sigma_c$}
because the $\Sigma_c\pi$ threshold is {well} below the
$\Lambda_c^*(2625)$ mass, {and} the convolution does
{not change the result much.}
{In the table,} we also show {the calculated decay widths
 of other assignments than $J^P=3/2^-$.}

We find that the assignment of $\lambda$-mode configuration with
$J^P=3/2^-$ for $\Lambda_c^*(2625)$ works very well to describe the 
small decay width of the $\Lambda_c^*(2625)\rightarrow\Sigma_c\pi$,
{while}
the assignment of $1/2^-$ {leads to} larger width than
the experimental value.  In contrast to the {case of}
$\Lambda_c^*(2595)(1/2^-)$, however, we cannot exclude the
possibilities of the $\rho$-mode configurations for
$\Lambda_c^*(2625)(3/2^-)$ {by the study of decay width,
{because the calculated $\Sigma_c\pi$ decay widths for $\lambda$-mode and two
$\rho$-modes with $J=3/2^-$ are accidentally similar to each other.
It is interesting, however, that these three modes give quite different
transition amplitudes for the $\Sigma_c^*(3/2^+)\pi$ decay as will be discussed later
in Sec.~\ref{sec:higher}, {although the $\Sigma_c^*\pi$ channel is
closed for $\Lambda_c^*(2625)$.} 
To
discuss the structure of 
$\Lambda_c^*(2625)$ in more detail, we need systematical analyses of the
mass spectrum~\cite{Yoshida:2015tia}, non-resonant three-body decay, and
so on.}


\begin{table*}[htb]
\caption{{Calculated decay widths of the
 $\Lambda_c^*(2765)\rightarrow\Sigma_c(2455)\pi$ and
 $\rightarrow\Sigma_c^*(2520)\pi$.  The quantum numbers of the
 $\lambda$- and $\rho$-modes are indicated by
 $(n_\lambda,\ell_\lambda)$ and $(n_\rho,\ell_\rho)$, and 
 $J_{\Lambda^*_{{c}}}(j)^P$ stands for the assigned spin and parity for $\Lambda_c^*$ with
the brown muck spin $j$.  
$[\Sigma_c^{(*)}\pi]^+$
 denotes the isospin summed width calculated by using the isospin
 average masses $M_{\Sigma_{{c}}}=2453.5$~(MeV), $M_{\Sigma^*_{{c}}}=2518.1$~(MeV),
 and $m_\pi=138.0$~(MeV). The ratio $R$ indicates the
 $\Sigma_c^*/\Sigma_c$ defined in the text.}
\label{tab:2765}}

\begin{center}
\begin{tabular}{cccccccc} 
\multicolumn{8}{l}{$\Lambda_c^*(2765)^+$ decay width ($M_{\Lambda^*}=2766.6$~(MeV))} \\\hline\hline
\multicolumn{3}{r}{decay channel} & full & $[\Sigma_c^{(*)}\pi]_{\rm total}$ & $[\Sigma_c\pi]^+$ & $[\Sigma_c^*\pi]^+$ & $R$ \\
\multicolumn{3}{r}{Experimental value $\Gamma_{\rm exp}$ (MeV)} & 50~\cite{Agashe:2014kda} &  & - & - & - \\
\multicolumn{3}{r}{momentum of final particle $q$ (MeV/c)} &  &  & 265 & 197 &  \\\hline\hline
 & $(n_\lambda,\ell_\lambda)$, $(n_\rho,\ell_\rho)$ & $J_\Lambda(j)^P$ &  &  &  &  &  \\\cline{2-8}
 & $(0,1)$, $(0,0)$ & $1/2(1)^-$ &  & 65.1--146.3  & 61.2--140.2  & 3.9--6.1  & 0.044--0.064  \\
 &  & $3/2(1)^-$ &  & 52.2--104.2  & 7.9--11.9  & 44.3--92.4  & 5.6--7.8  \\\cline{2-8}
 & $(0,0)$, $(0,1)$ & $1/2(0)^-$ &  & 0 & 0 & 0 & - \\\cline{3-8}
this work &  & $1/2(1)^-$ &  & 325.8--676.3  & 323.7--673.3  & 2.1--3.0  & 0.0044--0.0064 \\
$\Gamma$ &  & $3/2(1)^-$ &  & 210.4--413.5  & 4.2--5.8  & 206.2--407.7  & 49--70  \\\cline{3-8}
(MeV) &  & $3/2(2)^-$ &  & 9.4--13.1  & 7.6--10.5  & 1.9--2.7  & 0.25--0.26  \\
 &  & $5/2(2)^-$ &  & 6.3--8.8  & 3.4--4.7  & 2.9--4.2  & 0.87--0.90  \\\cline{2-8}
 & $(1,0)$, $(0,0)$ & $1/2(0)^+$ &  & 1.6--4.5  & 0.86--2.49  & 0.78--1.98  & 0.79--0.91 \\\cline{2-8}
 & $(0,2)$, $(0,0)$ & $3/2({2})^+$ &  & 4.7--10.9  & 4.4--10.1 & 0.33--0.72  & 0.071--0.076 \\
 &  & $5/2({2})^+$ &  & 1.9--4.4  & 0.13--0.32  & 1.77--4.04  & 12.8--13.8 \\
\end{tabular}
\end{center}

\end{table*}

\subsection{Decays of the higher excited $\Lambda_c^*$ baryons}
\label{sec:higher}
In Ref.~\cite{Agashe:2014kda}, 
{three more $\Lambda_c^*$ states are nominated,}
$\Lambda_c^*(2765)$, $\Lambda_c^*(2880)$, and $\Lambda_c^*(2940)$,
{though $\Sigma_c^*(2765)$ cannot be excluded {for $\Lambda_c^*(2765)$}.}
{Among them, spin of $\Lambda_c^*(2880)$ is the only quantum number
that is well determined {in experiment}.  The parity of
$\Lambda_c^*(2880)$ is assigned to be
positive, but it deserves being carefully examined.}
Therefore we consider {possible assignments of}
both {positive and negative} parity cases.
For these higher states, the $\Sigma_c^*(2520)\pi$ channel opens in
addition to the $\Sigma_c(2450)\pi$ channel.  
The ratio of
$\Gamma(\Sigma_c^*\pi)/\Gamma(\Sigma_c\pi)$ also can help {us} to determine the
quantum numbers, and the quark configuration as well.
In the following discussions, $\Sigma_c^{(*)}$ denotes
$\Sigma_c(2455)$ {with} $1/2^+$ or $\Sigma_c^*(2520)$ {with} $3/2^+$.

\subsubsection{{$\Lambda_c^*(2765)\rightarrow \Sigma_c^{(*)}\pi$
   decay}}
{The $\Lambda_c^*(2765)$ {baryon} is seen in $\Lambda_c^+\pi^+\pi^-$
channel as a broad peak~\cite{Agashe:2014kda,Artuso:2000xy}.
The width is reported as $\Gamma_{\rm exp}=50$~(MeV), but its quantum numbers are
still unknown.
For this baryon, we consider the $p$-wave excitations in $\lambda$- {or}
$\rho$-mode with  negative parity;
$\{(n_\lambda,\ell_\lambda),(n_\rho,\ell_\rho)\} =\{(0,1),(0,0)\}$ {or}
$\{(0,0),(0,1)\}$. {We also consider the possibility of} $s$-wave {or}
$d$-wave excitations in 
$\lambda$-mode with positive parity; 
${\{(n_\lambda,\ell_\lambda),(n_\rho,\ell_\rho)\}=}\{(1,0),(0,0)\}$ {or} $\{(0,2),(0,0)\}$.  
{Further studies on $\Lambda_c^*(2765)$ with other quark
configurations {are in progress and} will be discussed elsewhere~\cite{2765}.}}

{In Table~\ref{tab:2765}, we summarize the possible $\Lambda_c^*$ 
spin-parity {considered here} together with the calculated results.  Because
the partial decay widths are not measured yet, we show the isospin
summed width calculated by using the isospin-averaged masses
$M_{\Sigma^{(*)}_{{c}}}$ and $m_\pi$.  
The concrete forms of the helicity amplitudes are summarized in
Appendix.
We find that, for higher $j$, the decay width tends to be smaller due to
the suppression of the phase space for higher relative angular momentum
in the final state. 
}

\begin{table*}[hbt]
\caption{{Calculated decay width of the
 $\Lambda_c^*(2880)\rightarrow\Sigma_c(2455)\pi$ and
 $\rightarrow\Sigma_c^*(2520)\pi$.  
The quantum numbers of the $\lambda$- and $\rho$-modes are indicated by
$(n_\lambda,\ell_\lambda)$,$(n_\rho,\ell_\rho)$, and
 $J_{\Lambda^*_{{c}}}(j)^P$ stands for the assigned spin for $\Lambda_c^*$ with
the brown muck spin $j$ and the parity $P$.  For the $\{(0,1),(0,1)\}$
 configurations, we also show the total angular momentum
 $\vec{\ell}=\vec{\ell}_\lambda+\vec{\ell}_\rho$ as a subscript $\ell$
 in $J_{\Lambda^*_{{c}}}(j)^P_\ell$.
$[\Sigma_c^{(*)}\pi]^+$
 denotes the isospin summed width calculated by using the isospin
 average masses $M_{\Sigma_{{c}}}=2453.5$~(MeV), $M_{\Sigma^*_{{c}}}=2518.1$~(MeV),
 and $m_\pi=138.0$~(MeV). The ratio $R$ indicates the
 $\Sigma_c^*/\Sigma_c$ defined in the text.}
\label{tab:2880}}
\begin{center}
\begin{tabular}{cccccccc} 
\multicolumn{8}{l}{$\Lambda_c^*(2880)^+$ decay width  ($M_{\Lambda^*}=2881.53$~(MeV))} \\\hline\hline
\multicolumn{3}{r}{decay channel} & full & $[\Sigma_c^{(*)}\pi]_{\rm total}$ & $[\Sigma_c\pi]^+$ & $[\Sigma_c^*\pi]^+$ & $R$ \\
\multicolumn{3}{r}{Experimental value $\Gamma_{\rm exp}$ (MeV)} & $5.8\pm 1.1$~\cite{Agashe:2014kda} &  &  &  & 0.225~\cite{Abe:2006rz} \\
\multicolumn{3}{r}{momentum of final particle $q$ (MeV/c)} &  &  & 375 & 315 &  \\\hline\hline
 & $(n_\lambda,\ell_\lambda)$, $(n_\rho,\ell_\rho)$ & $J_\Lambda(j)^P$ &  &  &  &  &  \\\cline{2-8}
 & $(0,1)$, $(0,0)$ & $1/2(1)^-$ &  & 111.9--254.8  & 76.9--204.0  & 35.0--50.8  & 0.25--0.46 \\
 &  & $3/2(1)^-$ &  & 129.6--248.8  & 37.7--52.1  & 91.9--196.7  & 2.4--3.8  \\\cline{2-8}
 & $(0,0)$, $(0,1)$ & $1/2(0)^-$ &  & 0 & 0 & 0 & - \\\cline{3-8}
this work &  & $1/2(1)^-$ &  & 502.5--1129.7  & 483.9--1104.7  & 18.6--24.9  & 0.038--0.023  \\
$\Gamma$ &  & $3/2(1)^-$ &  & 439.3--919.5  & 20.0--25.6  & 419.3--893.9  & 21--35 \\\cline{3-8}
(MeV) &  & $3/2(2)^-$ &  & 52.8--68.5  & 36.0--46.0  & 16.7--22.4  & 0.46--0.49  \\
 &  & $5/2(2)^-$ &  & 42.0--55.3  & 16.0--20.5  & 26.0--34.9  & 1.6--1.7  \\\cline{2-8}
 & $(1,0)$, $(0,0)$ & $1/2(0)^+$ &  & 3.7--13.5  & 1.3--5.6  & 2.4--7.9  & 1.4--1.8 \\\cline{2-8}
 & $(0,2)$, $(0,0)$ & $3/2({2})^+$ &  & 16.3--39.5  & 13.9--34.2  & 2.4--5.3  & 0.16--0.17 \\
 &  & $5/2({2})^+$ &  & 11.2--26.1  & 1.2--2.8  & 9.9--23.3  & 8.1--8.4 \\\cline{2-8}
 & $(0,0)$, $(1,0)$ & $1/2(0)^+$ &  & 16.5--40.2  & 7.0--18.2  & 9.5--22.1  & 1.2--1.4 \\\cline{2-8}
 & $(0,0)$, $(0,2)$ & $3/2({2})^+$ &  & 44.8--85.4  & 39.5--76.0  & 5.3--9.4  & 0.12--0.13  \\
 &  & $5/2({2})^+$ &  & 27.8--52.2  & 1.4--2.6  & 26.4--49.5  & 18.7--18.9 \\\cline{2-8}
 &  &  &  &  &  &  &  \\
 & $(n_\lambda,\ell_\lambda)$, $(n_\rho,\ell_\rho)$ & $J_\Lambda(j)_\ell^P$ &  &  &  &  &  \\\cline{2-8}
 & $(0,1)$, $(0,1)$ & $5/2(2)_2^+$ &  & 51.7--109.6  & 1.8--3.5  & 49.9--106.1  & 27.5--30.1 \\
 &  & $5/2(2)_1^+$ &  & 0.63--1.68  & 0 & 0.63--1.68  & $(\infty)$ \\
 &  & $5/2(3)_2^+$ &  & 2.9--5.8  & 2.1--4.0  & 0.85--1.73  & 0.41--0.43  \\
\end{tabular}
\end{center}

\end{table*}


{In the last column in Table~\ref{tab:2765}, we also show the ratio of
{the decay widths to $\Sigma_{c}(2455)\pi$ and
$\Sigma_{c}^{\ast}(2520)\pi$} 
defined by
\begin{equation}
 R=\frac{\Gamma(\Lambda_c^*\rightarrow
  \Sigma_c^*\pi)}{\Gamma(\Lambda_c^*\rightarrow \Sigma_c\pi)}\ .
\label{eq:R}
\end{equation}
We} find the order of magnitudes of the ratio $R$ are quite different
for different configurations even if the spin-parity is the same,
{e.g. ${J_{\Lambda_{c}^{\ast}}(j)^{P}=}3/2(1)^-(\lambda\text{-mode})$,
$3/2(1)^-(\rho\text{-mode})$ and $3/2(2)^-(\rho\text{-mode})$.} 
{In fact, these three modes give the similar widths for the
$\Sigma_c\pi$ decay as discussed in the previous section, but give quite
different widths for $\Sigma_c^*\pi$.}
In principle, the $\Lambda_c^*(3/2^-)$ baryon {can} decay by $s$-wave to
$\Sigma_c^*(3/2^+) \pi(0^-)$, while {it} decays by
$d$-wave to $\Sigma_c(1/2^+) \pi(0^-)$.  {Then the ratio $R$ can be
expressed by
\begin{equation}
 R=\frac{\Gamma(\Sigma_c^*\pi)_s+\Gamma(\Sigma_c^*\pi)_d}{\Gamma(\Sigma_c\pi)_d},
\end{equation}
which is, in general, larger than unity.
This is the case for the $J_{{\Lambda_{c}^{\ast}}}(j)^P=3/2(1)^-$ as
\begin{eqnarray}
   R(3/2(1)^-(\lambda\text{-mode})) &=& 5.6 \text{--} 7.8 \ ,\\
   R(3/2(1)^-(\rho\text{-mode})) &=& 49 \text{--} 70 \ .
\end{eqnarray}
In contrast,} the brown muck $j^P=2^-$ state cannot decay by $s$-wave
{to the brown muck $1^+$ state in} $\Sigma_c^*(3/2^+) \pi(0^-)$
because of the spin-parity conservation. 
{This is another example of the selection rules in the heavy quark limit. 
Due to the absence of $s$-wave contribution,}
the ratio $R$ is smaller than unity for $3/2(2)^-$ as
\begin{equation}
  R(3/2(2)^-(\rho\text{-mode})) =
   \frac{\Gamma(\Sigma_c^*\pi)_d}{\Gamma(\Sigma_c\pi)_d} =0.25 \text{--}
   0.26 \ .
\end{equation}
{In this configuration, the amplitudes of $\Sigma_c\pi$ and
$\Sigma_c^*\pi$ decays} are
the same except {the momentum $q$ of pion} as discussed in
Ref.~\cite{Isgur:1991wq}.
%
%
Here, we stress that the $s$-wave suppression {for $J_{{\Lambda_{c}^{\ast}}}^P=3/2^-$}
is found only in the {case of} $j^P=2^-$, and not in the other quark
configurations. 
This is the same phenomenon that the $1/2(0)^-$ state cannot decay into
$\Sigma_c^{(*)}\pi$ as mentioned in Sec.~\ref{sec:brown}, and also is seen
{for the decay of}
the $\Lambda_c^*(2880)$ as discussed in the next section.

{As for the magnitude of the decay width, we find that the
assignments of $J_{{\Lambda_{c}^{\ast}}}(j)^P=1/2(1)^-$ and $3/2(1)^-$ ($\ell_\rho=1$)
give rather large decay widths due to the $s$-wave nature into either
$\Sigma_c(1/2^+)\pi$ or $\Sigma_c^*(3/2^+)\pi$. We can exclude these assignments
because the resulting decay widths are too large.}
Calculated widths for $\lambda$-modes ($\ell_\lambda = 1$) are slightly
larger as compared with the observed full width,  
which does not seem inconsistent if we consider the uncertainty of
$g_A^q$ . However, by taking into account contributions of decays into
non-resonant three-body $\Lambda \pi \pi$, these $\lambda$-mode states
will receive a larger full width, with which the possibility for them to
be identified with $\Lambda_c^*(2765)$ might decrease.

{Among the considered assignments in this article, the other
assignments $J_{{\Lambda_{c}^{\ast}}}(j)^P=1/2(0)^-$, $1/2(0)^+$, $3/2(2)^-$,
$3/2(2)^+$, $5/2(2)^-$ and $5/2(2)^+$ cannot be excluded because the
total $\Sigma_c^{(*)}\pi$ decay width is consistent with the experimental
value. The ratio $R$, however, takes different value reflecting the
structure of the $\Lambda_c^*$ baryon which will help to determine
the quantum numbers.

\subsubsection{{$\Lambda_c^*(2880)\rightarrow \Sigma_c^{(*)}\pi$
   decay}}

The $\Lambda_c(2880)$ charmed baryon is observed in
$\Lambda_c\pi\pi$ channel~\cite{Abe:2006rz,Artuso:2000xy} as well as in
$pD^0$ channel~\cite{Aubert:2006sp}.  The spin is determined as $5/2$
from the angular distribution of the decay into
$\Sigma_c(2455)\pi$~\cite{Abe:2006rz}.  In PDG~\cite{Agashe:2014kda},
the parity is {assigned to be}
positive {from the analysis of}
$\Sigma_c^*/\Sigma_c$
branching ratio {in comparison} with the prediction of the chiral
perturbation~\cite{Cheng:2006dk} 
with the heavy quark symmetry~\cite{Isgur:1991wq}.
{However, as discussed in \cite{Cheng:2006dk} {a} subtlety arises when
calculating the ratio.}


In Table~\ref{tab:2880} we summarize the quark configurations considered
here for $\Lambda_c^*(2880)$.  By comparing the observed full width
$\Gamma_{\rm exp}=5.8$~MeV and calculated total one-pion decay width, we can
exclude all of the $p$-wave configurations with the negative parity including
$5/2^-$. As for $5/2^-$ with $\rho$-mode excitation, 
{both of the decays
$\Lambda_c^*(5/2^-)\rightarrow\Sigma_c(1/2^+)\pi$ and
${\Lambda_c^*(5/2^-)\rightarrow}\Sigma_c^*(3/2^+)\pi$ {($(j^P=2^-)\rightarrow(j^P=1)^+0^-$ in
terms of the brown muck)} go through by $d$-wave, and}
the
$\Sigma_c^*/\Sigma_c$ ratio $R$ {is larger than unity as,
\begin{equation}
 R(5/2(2)^-;\rho) = 1.6 \text{--}1.8, \
\end{equation}
which does not agree with the experimental value
$R=0.225\pm 0.062 \pm 0.010$~\cite{Abe:2006rz}.}
This conclusion is consistent with the chiral perturbation 
calculation with 
heavy quark symmetry~\cite{Cheng:2006dk,Isgur:1991wq}.

\begin{table*}[hbt]
\caption{{Calculated decay width of the
 $\Lambda_c^*(2940)\rightarrow\Sigma_c(2455)\pi$ and
 $\rightarrow\Sigma_c^*(2520)\pi$.  
The quantum numbers of the $\lambda$- and $\rho$-modes are indicated by
$(n_\lambda,\ell_\lambda)$,$(n_\rho,\ell_\rho)$, and
 $J_{\Lambda^*_{{c}}}(j)^P$ stands for the assigned spin for $\Lambda_c^*$ with
the brown muck spin $j$ and the parity $P$.  For the $\{(0,1),(0,1)\}$
 configurations, we also show the total angular momentum
 $\vec{\ell}=\vec{\ell}_\lambda+\vec{\ell}_\rho$ as a subscript $\ell$
 in $J_{\Lambda^*_{{c}}}(j)^P_\ell$.
$[\Sigma_c^{(*)}\pi]^+$
 denotes the isospin summed width calculated by using the isospin
 average masses $M_{\Sigma_{{c}}}=2453.5$~(MeV), $M_{\Sigma^*_{{c}}}=2518.1$~(MeV),
 and $m_\pi=138.0$~(MeV). The ratio $R$ indicates the
 $\Sigma_c^*/\Sigma_c$ defined in the text.}
\label{tab:2940}}
\begin{center}
\begin{tabular}{cccccccc} 
\multicolumn{8}{l}{$\Lambda_c^*(2940)^+$ decay width ($M_{\Lambda^*}=2939.3$~(MeV))} \\\hline\hline
\multicolumn{3}{r}{decay channel} & full & $[\Sigma_c^{(*)}\pi]_{\rm total}$ & $[\Sigma_c\pi]^+$ & $[\Sigma_c^*\pi]^+$ & $R$ \\
\multicolumn{3}{r}{Experimental value $\Gamma$ (MeV)} & $17^{+8}_{-6}$~\cite{Agashe:2014kda} &  & (seen) & - &  \\
\multicolumn{3}{r}{momentum of final particle $q$ (MeV/c)} &  &  & 427 & 369 &  \\\hline\hline
 & $(n_\lambda,\ell_\lambda)$, $(n_\rho,\ell_\rho)$ & $J_\Lambda(j)^P$ &  &  &  &  &  \\\cline{2-8}
 & $(0,1)$, $(0,0)$ & $1/2(1)^-$ &  & 144.8--313.8  & 73.8--215.4  & 71.0--98.4  & 0.46--0.96 \\
 &  & $3/2(1)^-$ &  & 182.2--332.0  & 65.4--85.7  & 116.8--246.3  & 1.8--2.9  \\\cline{2-8}
 & $(0,0)$, $(0,1)$ & $1/2(0)^-$ &  &  &  &  &  \\\cline{3-8}
this work &  & $1/2(1)^-$ &  & 557.0--1299.3  & 519.3--1250.9  & 37.6--48.3  & 0.039--0.072 \\
$\Gamma$ &  & $3/2(1)^-$ &  & 536.5--1152.9  & 34.6--42.2  & 501.8--1110.7  & 15--26  \\\cline{3-8}
(MeV) &  & $3/2(2)^-$ &  & 96.2--119.4  & 62.3--75.9  & 33.9--43.5  & 0.54--0.57  \\
 &  & $5/2(2)^-$ &  & 80.4--101.4  & 27.7--33.7  & 52.7--67.7  & 1.9--2.0  \\\cline{2-8}
 & $(1,0)$, $(0,0)$ & $1/2(0)^+$ &  & 3.7--17.4  & 1.1--6.4  & 2.7--11.0  & 1.7--2.5 \\\cline{2-8}
 & $(0,2)$, $(0,0)$ & $3/2({2})^+$ &  & 24.9--61.7  & 20.1--51.0  & 4.8--10.8  & 0.21--0.24 \\
 &  & $5/2({2})^+$ &  & 19.8--46.6  & 2.8--5.9  & 17.1--40.7  & 6.2--6.9  \\\cline{2-8}
 &  &  &  &  &  &  &  \\
 & $(n_\lambda,\ell_\lambda)$, $(n_\rho,\ell_\rho)$ & $J_\Lambda(j)_\ell^P$ &  &  &  &  &  \\\cline{2-8}
 & $(0,1)$, $(0,1)$ & $7/2(3)_2^+$ &  & 5.8--11.1 & 2.6--4.8  & 3.2--6.2 & 1.22--1.29 \\
\end{tabular}
\end{center}

\end{table*}

For the spin-parity $5/2^+$ case, we can consider five
configurations as shown in Table~\ref{tab:2880}; one $d$-wave excitation
in $\lambda$-motion ({$5/2(2)^+$ with} $\ell_\lambda=2$, denoted by $\lambda\lambda$), the one in $\rho$-motion
({$5/2(2)^+$ with} $\ell_\rho=2$, denoted by $\rho\rho$), and three {double-$p$-wave 
excitations in $\lambda$- and $\rho$-{motions}
{($J_\Lambda(j)_\ell^P=5/2(2)_1^+$, $5/2(2)_2^+$, $5/2(3)_2^+$
where $\vec{\ell}=\vec{\ell}_\ell+\vec{\ell}_\rho$ with
$(\ell_\lambda,\ell_\rho)=(1,1)$, denoted by $\lambda\rho$).} 
{Some of these configurations give consistent decay width with}
the observed full width $\Gamma_{\rm exp}=5.8$~(MeV).
{As for the $\Sigma_c^*/\Sigma_c$ ratio, {however,} we obtain considerably
different values for different configurations like,}
\begin{equation}
\begin{split}
  R(5/2({2})^+;{\lambda}\lambda)&=8.1\text{--}8.4\ , \\
 R(5/2({2})^+;{\rho}\rho)&=18.7\text{--}18.9\ , \\
 R(5/2(2)_2^+;\lambda\rho)&=27.5\text{--}30.1\ , \\
 R(5/2(2)_1^+;\lambda\rho)&=(\infty) \ , \\
 R(5/2(3)_2^+;\lambda\rho)&=0.41\text{--}0.43 \ ,
\end{split}
\end{equation}
{where the ambiguities of model parameters are almost canceled.}
Note that $(\infty)$ for $5/2(2)_1^+(\lambda\rho)$ state is due to
the {zero} decay width into $\Sigma_c\pi$.
{Among} these five configurations, we find that {\em only one}
configuration {$(5/2(3)_2^+;\lambda\rho)$} with the brown muck spin $j=3$
with $\ell=2$ agrees  
both with the small ratio $R<1$ and with the magnitude of total decay
width.
{This seems to contrast with the calculation in Ref.~\cite{Cheng:2006dk},
where the other quark configuration for $5/2^+$ also gives the small
$R$.}

{This discrepancy can be explained as follows.}
{The decay of
$\Lambda_c^*(5/2^+) \rightarrow \Sigma_c(1/2^+) \pi$
goes through only by the $f$-wave in the final two-body state, while
the decay $\Lambda_c^*(5/2^+) \rightarrow \Sigma_c^*(3/2^+) \pi$ can go
through both by $f$ and $p$-waves.   
The discussion based on the heavy quark limit leading to the model
independent relation is possible only when the same $f$-waves are
{taken}, which is completely contaminated by the presence of the
$p$-wave contribution. As shown explicitly in Appendix, the amplitude
for $\Lambda_c^*(5/2^+) \rightarrow \Sigma_c^*(3/2^+) \pi$
can contain the $p$-wave contribution ($c_1$ term {in Eq.~(\ref{eq:sd})}).}
{Thus we have}
\begin{equation}
 R{(5/2^+)}=\frac{\Gamma(\Sigma_c^*\pi)_p+\Gamma(\Sigma_c^*\pi)_d}
{\Gamma(\Sigma_c\pi)_d} > 1,
\end{equation}
except the case of $5/2(3)_2^+$.}
Only for {the case of} $5/2(3)_2^+$, {$p$-wave contribution
($\tilde{c}_1$ term in Eq.~(\ref{eq:pp}))} is zero
because of the conservation of the brown muck spin-parity; the brown muck of
$3^+$ cannot decay into $1^+$ with the pion $0^-$ in $p$-wave, {which
leads to
\begin{equation}
 R(5/2(3)_2^+;\lambda\rho)=\frac{\Gamma(\Sigma_c^*\pi)_d}{\Gamma(\Sigma_c\pi)_d}
  < 1 \ .
\end{equation}
We stress here again that the $p$-wave suppression is found only in the
case of $5/2(3)^+$ with $\ell=2$, and not in the other states with $5/2^+$.}
Here it is worth to mention that the ${\cal O}(q^1)$
contribution, which allows us to distinguish the possible quark
configurations for the same spin-parity, appears only in
$A^{\nabla\cdot\sigma}$ term arising from the 
axial-vector coupling $\gamma_\mu\gamma_5$ of the pion.

{If $\Lambda_c^*(2880)$ is assigned as a $\lambda\rho$-mode state,
a question arises where the $\lambda\lambda$-mode states with
$\ell_\lambda=2$ are.
Excitation energies of the $\lambda\lambda$-mode states are expected to be
lower than those of the $\lambda\rho$-mode states. 
Other information such as production rates as discussed in
Ref.~\cite{Kim:2014qha} is helpful {to solve this problem}, for which an
experimental measurement is planned in J-PARC~\cite{E50}.}

\subsubsection{{$\Lambda_c(2940)\rightarrow\Sigma_c^{(*)}\pi$
   decay}}
{
As for $\Lambda_c^*(2940)$,  {a} narrow peak is observed both in $pD^0$
channel~\cite{Aubert:2006sp} and in $\Sigma_c\pi$
channel~\cite{Abe:2006rz}.  {The total width is
$\Gamma_{\rm exp}=17^{+8}_{-6}$~(MeV)~\cite{Agashe:2014kda}.  The
spin-parity is not {determined}.}

{In Table~\ref{tab:2940}, we show the calculated one-pion decay
widths together with the considered quark configurations for
$\Lambda_c^*(2940)$.  In the previous section, we pointed
out the possibility that $\Lambda_c^*(2880)$ is $5/2(3)_2^+$ excitation.
{If this is the case,
a new question arises;} which $Y_c$ baryon is the 
{partner of the {HQS} doublet} {possessing}
$7/2(3)_2^+$.  To discuss the possibility of
$\Lambda_c^*(2940)$ being the doublet {partner} {of $\Lambda_c^*(2880)$}, we also show
the one-pion decay width with the $7/2(3)_2^+$ assignment for
$\Lambda_c^*(2940)$ in the last line of Table~\ref{tab:2940}.
We can see that this assignment can be consistent {with} the experimental
full width in \cite{Agashe:2014kda} in the sense that the calculated
total one-pion emission decay width does not exceed the reported full
width.  {For} the same reason, the negative parity assignments can
be excluded for the $\Lambda_c^*(2940)$.  Similarly to other
$\Lambda_c^*$ baryons, the partial decay widths and/or the 
$\Sigma_c^*/\Sigma_c$ ratio
will help to determine the quantum numbers
and the possible quark configuration as well.
}


\section{Summary}
\label{sec:summary}
{
We have systematically evaluated the decay widths 
of the charmed baryons $\Lambda_c^*(2595)$, $\Lambda_c^*(2625)$,
$\Lambda_c^*(2765)$, $\Lambda_c^*(2880)$, and $\Lambda_c^*(2940)$ 
into {$\Sigma_c(2455)\pi$ and $\Sigma_c^*(2520)\pi$}, as well as $\Sigma_c(2455)$ and
$\Sigma_c^*(2520)$ into $\Lambda_c\pi$
within the non-relativistic quark model.
{We have emphasized the usefulness of working in the baryon wave
functions constructed {to be consistent with} heavy quark symmetry. 
This provides various selection rules associated with the pion emission
between brown muck of the baryons.}
Our findings are as follows:
\begin{itemize}
 \item For the low-lying $\Lambda_c^*(2595)$ and $\Lambda_c^*(2625)$ baryons
       the quark model descriptions as the $\lambda$-mode excitations with
       spin-0 diquark can explain the decay properties very well.

 \item The derivative coupling derived from the axial-vector interaction
       of $\pi qq$ is {essentially important} to produce the experimental decay rate
       of $\Lambda_c^*(2595)$.

 \item Only one quark configuration {$J_{\Lambda_{c}^{\ast}}(j)^{P}=5/2(3)^{+}_{2}$} for $\Lambda_c^*(2880)$ among the possible
       five $5/2^+$ configurations can lead to the
       consistent result with the experimental data,
       while all other four configurations of $5/2^+$ cannot if the
       $p$-wave is properly considered.  We note that the HQS does not
       necessarily lead to the small decay ratio of
       $\Gamma(\Sigma_c^*\pi)/(\Sigma_c\pi)$ for $5/2^+$.
       This fact calls an attention to the discussion based on the
       HQS~\cite{Isgur:1991wq,Cheng:2006dk} which
       requires decays in only one partial wave. 

 \item Having the above conclusion, we have discussed the
       possibility of $\Lambda_c^*(2940)$ being the HQS doublet partner of 
       $\Lambda_c^*(2880)$ possessing $7/2(3)_2^+$.  Here we emphasize
       that our results concerning the possible HQS doublet,
       $\Lambda_c^*(2880)$ and $\Lambda_c^*(2940)$, can be reached with
       $jj$ coupling scheme which respects the heavy quark symmetry.

 \item The ratios of
       $\Gamma(\Sigma_c^*\pi)/\Gamma(\Sigma_c\pi)$ are
       considerably different for different quark configurations
       even if the baryon spin-parity is the same.
       {This fact is particularly useful to know the structure of
       charmed baryons.} 
\end{itemize}
}


{
In this study, we have discussed the various constraints for the
one-pion emission decays 
due to the selection rules associated with the brown muck spin $j$
conservation.
We have to keep it in mind, however, that there is still small breaking
of heavy quark symmetry for charm quark. 
The study along this line will be left for future works.}}

{In our discussions in the quark model, we have considered only the
excitations of valence quarks. We expect that they provide a good
description for low lying states.  For higher excitations, however,
there {may be} {other modes such as pair creations of quark and
anti-quark, gluon excitations and so on. The former can be}
taken into account in the quark model by couplings to mesons
or by an unquenched configurations~\cite{Bijker:2012zza}, and in
effective hadron models by hadronic molecule
configurations~\cite{Mizutani:2006vq,Haidenbauer:2010ch}. 
{The present systematic studies will help us to know w}here and how
these configurations beyond the quark model ones show up {{which}}
should be studied in the future J-PARC experiments.
}

\section*{acknowledgment}
{The authors are grateful to K.~Tanida and T.~Yoshida for various
discussions.  This work is supported by Grants-in-Aid for Scientific
Research (Grants No. JP26400275(C) for H.~Nagahiro), 
{(Grants No. JP15K17641 for S.~Y.)}, (Grants No. JP26400273(C) for
A.~H.), (Grants No. JP16H02188(A) for H.~Noumi) {and (Grants No. JP25247036(A) for
M.~O., S.~Y. and A.~H.)}. } 

\appendix
\section{Harmonic oscillator wave functions}
\label{sec:App_R}
The radial functions $R_{n\ell}(\zeta)$ are given as,
\begin{eqnarray} 
 R_{00}(\zeta)&=&\frac{2a_\zeta^{3/2}}{\pi^{1/4}} e^{-a_\zeta^2\zeta^2/2}, \\
 R_{01}(\zeta)&=&\left(\frac{8}{3}\right)^{1/2}\frac{a_\zeta^{5/2}}{\pi^{1/4}}\zeta,
  e^{-a_\zeta^2\zeta^2/2}, \\
 R_{02}(\zeta)&=&\left(\frac{16}{15}\right)^{1/2}\frac{a_\zeta^{7/2}}{\pi^{1/4}}\zeta^2,
  e^{-a_\zeta^2\zeta^2/2}, \\
 R_{10}(\zeta)&=&\frac{\sqrt{6}a_\zeta^{7/2}}{\pi^{1/4}}\left(1-\frac{2}{3}a_\zeta^2\zeta^2\right)
  e^{-a_\zeta^2\zeta^2/2},
\end{eqnarray}
where
\begin{equation}
 a_\zeta=\sqrt{m_\zeta\omega_\zeta}.
\end{equation}
The $\zeta$ is either $\lambda$ or $\rho$.  The reduced masses of
$m_\lambda$ and $m_\rho$ are defined in Eq.~(\ref{eq:reducedmass}).  

\section{Matrix elements}
\label{sec:App_M}
In this Appendix, we summerize the concrete forms of the helicity
amplitudes $A_h$.  
\subsection{Ground state $\Sigma_c$ decays}
The amplitudes for the decays of
$\Sigma_c^{(*)}\rightarrow\Lambda_c(1/2^+)\pi^-$ are given by
\begin{equation}
 \begin{split}
-iA_{1/2}^{\nabla\cdot\sigma}&=G\dfrac{\omega_\pi}{m} c
\left(\dfrac{1}{2}q_\lambda+q_\rho\right)F(q),\\
 iA_{1/2}^{q\cdot\sigma}&=-G\dfrac{q}{m}
c
\left(\dfrac{M}{2m+M}\omega_\pi-2m\right)F(q),
 \end{split} 
\end{equation}
where the coefficent $c$ is given as
\begin{equation}
 c=
\begin{cases}
-1/\sqrt{3} & \text{ for } \Sigma_c(1/2^+) \ ,\\
\sqrt{2/3}  & \text{ for } \Sigma_c^*(3/2^+)\ . 
\end{cases}
\end{equation}
The factor $G$ denotes the coupling constant and the normalizations as,
\begin{equation}
 G = \frac{g_A^q}{2f_\pi}\sqrt{2M_{\Lambda_{{c}}}}\sqrt{2M_{\Sigma^{(\ast)}_{{c}}}} \ ,
\end{equation}
and the function $F(q)$ denotes the gaussian form factor as
\begin{equation}
 F(q)=e^{-{q_\lambda^2}/{4a_\lambda^2}}e^{-{q_\rho^2}/{4a_\rho^2}}\ .
\end{equation}

\subsection{Negative parity $\Lambda_c^*(J^-)$ decays}
\label{sec:Ah}
The amplitudes for the decays of the negative parity excitations with
$p$-wave of $\Lambda_c^*(J^-)\rightarrow \Sigma_c^{(*)}\pi$ are given by
\begin{equation}
 \begin{split}
-iA_h^{\nabla\cdot\sigma}&=iG \dfrac{\omega_\pi}{m} \
\left\{
c_0 a_\zeta + c_2 \left(\dfrac{1}{2}q_\lambda+q_\rho\right)\dfrac{q_\zeta}{a_\zeta}
\right\}F(q) \ ,
 \\  
-iA_h^{q\cdot\sigma}&=iG\dfrac{q}{m}\left(\dfrac{M}{2m+M}\omega_\pi-2m\right)(-1)
c_2\dfrac{q_\zeta}{a_\zeta} F(q) \ ,
 \end{split}
\label{eq:p}
\end{equation}
where the coefficients $c_0$ and $c_2$ are summarized in
Table~\ref{tab:p}.  The subscript $\zeta$ is either $\lambda$ or $\rho$,
depending on the $\lambda$- or $\rho$-mode excitations.

\begin{table}[tb]
\caption{Coefficients for the negative partiy $\Lambda_c^*$ decays in Eq.~(\ref{eq:p})
\label{tab:p}.}
\begin{center}
\begin{tabular}{cccccc} 
\multicolumn{6}{l}{$\lambda$-mode excitation ($\zeta=\lambda$)} \\
$(n_\lambda,\ell_\lambda)$ $(n_\rho,\ell_\rho)$ & $J_\Lambda(j)^P$ & $J_\Sigma^P$ & $h$ & $c_0$ & $c_2$ \\\hline\hline
$(0,1)$,  $(0,0)$ & $1/2(1)^-$ & $1/2^+$ & 1/2 & $-\frac{1}{\sqrt{2}}$ & $\frac{1}{3\sqrt{2}}$ \\
 &  & $3/2^+$ & 1/2 & 0 & $-\frac{1}{3}$ \\\cline{2-6}
 & $3/2(1)^-$ & $1/2^+$ & 1/2 & 0 & $-\frac{1}{3}$ \\
 &  & $3/2^+$ & 1/2 & $-\frac{1}{\sqrt{2}}$ & $\frac{\sqrt{2}}{3}$ \\
 &  &  & 3/2 & $-\frac{1}{\sqrt{2}}$ & 0 \\\hline
 &  & \multicolumn{1}{c}{} & \multicolumn{1}{c}{} &  &  \\
\multicolumn{6}{l}{$\rho$-mode exciation ($\zeta=\rho$)} \\\hline\hline
$(0,0)$,  $(0,1)$ & $1/2(0)^-$ & $1/2^+$ & 1/2 & 0 & 0 \\
 &  & $3/2^+$ & 1/2 & 0 & 0 \\\cline{2-6}
 & $1/2(1)^-$ & $1/2^+$ & 1/2 & 2 & $-\frac{1}{3}$ \\
 &  & $3/2^+$ & 1/2 & 0 & $-\frac{1}{3\sqrt{2}}$ \\\cline{2-6}
 & $3/2(1)^-$ & $1/2^+$ & 1/2 & 0 & $-\frac{1}{3\sqrt{2}}$ \\
 &  & $3/2^+$ & 1/2 & 2 & $-\frac{1}{6}$ \\
 &  &  & 3/2 & 2 & $-\frac{1}{2}$ \\\cline{2-6}
 & $3/2(2)^-$ & $1/2^+$ & 1/2 & 0 & $\frac{1}{\sqrt{10}}$ \\
 &  & $3/2^+$ & 1/2 & 0 & $\frac{1}{2\sqrt{5}}$ \\
 &  &  & 3/2 & 0 & $-\frac{1}{2\sqrt{5}}$ \\\cline{2-6}
 & $5/2(2)^-$ & $1/2^+$ & 1/2 & 0 & $\frac{1}{\sqrt{15}}$ \\
 &  & $3/2^+$ & 1/2 & 0 & $\frac{1}{\sqrt{30}}$ \\
 &  &  & 3/2 & 0 & $\frac{1}{\sqrt{5}}$ \\\hline
\end{tabular}
\end{center}
\end{table}

\subsection{Positive parity $\Lambda_c^*(J^+)$ decays}

\begin{table}
\caption{Coefficients for the positive partiy $\Lambda_c^*(J^+)$ decays with
$s$-wave ($n_\zeta=1$) or $d$-wave ($\ell_\zeta=2$) in
 Eq.~(\ref{eq:sd}).
\label{tab:sd}}
 \begin{center}
\begin{tabular}{cccccc} 
\multicolumn{6}{l}{$\lambda$-mode excitation ($\zeta=\lambda$)} \\
$(n_\lambda,\ell_\lambda)$ $(n_\rho,\ell_\rho)$ & $J_{\Lambda^{\ast}_{{c}}}(j)^P$ & $J_{\Sigma^{{(\ast)}}_{{c}}}$ & $h$ & $c_1$ & $c_3$ \\\hline\hline
$(1,0)$, $(0,0)$ & $1/2(0)^+$ & $1/2^+$ & 1/2 & $\frac{1}{3\sqrt{2}}$ & $-\frac{1}{6\sqrt{2}}$ \\
 &  & $3/2^+$ & 1/2 & $-\frac{1}{3}$ & $\frac{1}{6}$ \\\hline
$(0,2)$,  $(0,0)$ & $3/2(2)^+$ & $1/2^+$ & 1/2 & $\frac{1}{3}\sqrt{\frac{5}{2}}$ & $-\frac{1}{3\sqrt{10}}$ \\
 &  & $3/2^+$ & 1/2 & $-\frac{1}{6\sqrt{5}}$ & $\frac{1}{3\sqrt{5}}$ \\
 &  &  & 3/2 & $-\frac{1}{2\sqrt{5}}$ & 0 \\\cline{2-6}
 & $5/2(2)^+$ & $1/2^+$ & 1/2 & 0 & $\frac{1}{2\sqrt{15}}$ \\
 &  & $3/2^+$ & 1/2 & $\sqrt{\frac{3}{10}}$ & $-\frac{1}{\sqrt{30}}$ \\
 &  &  & 3/2 & $\frac{1}{\sqrt{5}}$ & 0 \\
\multicolumn{6}{l}{$\rho$-mode excitation ($\zeta=\rho$)} \\\hline\hline
$(0,0)$, $(1,0)$ & $1/2(0)^+$ & $1/2^+$ & 1/2 & $\frac{\sqrt{2}}{3}$ & $-\frac{1}{6\sqrt{2}}$ \\
 &  & $3/2^+$ & 1/2 & $-\frac{2}{3}$ & $\frac{1}{6}$ \\\hline
$(0,0)$,  $(0,2)$ & $3/2(2)^+$ & $1/2^+$ & 1/2 & $\frac{\sqrt{10}}{3}$ & $-\frac{1}{3\sqrt{10}}$ \\
 &  & $3/2^+$ & 1/2 & $-\frac{1}{3\sqrt{5}}$ & $\frac{1}{3\sqrt{5}}$ \\
 &  &  & 3/2 & $-\frac{1}{\sqrt{5}}$ & 0 \\\cline{2-6}
 & $5/2(2)^+$ & $1/2^+$ & 1/2 & 0 & $\frac{1}{2\sqrt{15}}$ \\
 &  & $3/2^+$ & 1/2 & $\sqrt{\frac{6}{5}}$ & $-\frac{1}{\sqrt{30}}$ \\
 &  &  & 3/2 & $\frac{2}{\sqrt{5}}$ & 0 \\\hline
\end{tabular}
\end{center}
\end{table}

The amplitudes for the decays of the positive parity excitations with
$s$-wave ($n_\zeta=1$) or $d$-wave ($\ell_\zeta=2$) of
$\Lambda_c^*(J^-)\rightarrow \Sigma_c^{(*)}\pi$ 
are given by 
\begin{equation}
\begin{split}
  -iA_h^{\nabla\cdot\sigma}&=G\dfrac{\omega_\pi}{m}
\left\{
c_1 q_\zeta + c_3 \left(\dfrac{1}{2}q_\lambda+q_\rho\right)
\dfrac{q_\zeta^2}{a_\zeta^2}
\right\}F(q), \\
-iA_h^{q\cdot\sigma}&=G\dfrac{q}{m}
\left(\dfrac{M}{2m+M}\omega_\pi-2m\right)(-1)c_3\dfrac{q_\zeta^2}{a_\zeta^2}
  F(q).
\end{split}
\label{eq:sd}
\end{equation}
where the coefficients $c_1$ and $c_3$ are summarized in
Table~\ref{tab:sd}.  The subscript $\zeta$ is either $\lambda$ or $\rho$,
depending on the $\lambda$- or $\rho$-mode excitations.

\begin{table}
\caption{Coefficients for the positive partiy $\Lambda_c^*(J^+)$ decays with
$\lambda$-$\rho$ mixed excitations in
 Eq.~(\ref{eq:pp}). {$\ell$ denotes the total angular momentum
 defined by $\vec{\ell}=\vec{\ell}_\lambda+\vec{\ell}_\rho$.}
\label{tab:pp}}
 \begin{center}
\begin{tabular}{ccccccc} 
\multicolumn{7}{l}{$\lambda$-$\rho$ mixed excitation} \\
$(n_\lambda,\ell_\lambda)$ $(n_\rho,\ell_\rho)$ & $J_{\Lambda^{\ast}_{c}}(j)^P$ & $\ell$ & $J_{\Sigma_{c}^{(\ast)}}^P$ & $h$ & $\tilde{c}_1$ & $\tilde{c}_3$ \\\hline\hline
$(0,1)$,  $(0,1)$ & $5/2(2)^+$ & 2 & $1/2^+$ & 1/2 & 0 & $\frac{1}{3}\sqrt{\frac{1}{5}}$ \\
 &  &  & $3/2^+$ & 1/2 & $-\frac{3}{2}\sqrt{\frac{1}{10}}$ & $\frac{1}{3}\sqrt{\frac{1}{10}}$ \\
 &  &  & $3/2^+$ & 3/2 & $-\frac{3}{2}\sqrt{\frac{1}{15}}$ & $\frac{1}{\sqrt{15}}$ \\\cline{3-7}
 &  & 1 & $1/2^+$ & 1/2 & 0 & 0 \\
 &  &  & $3/2^+$ & 1/2 & $-\frac{1}{2}\sqrt{\frac{3}{10}}$ & 0 \\
 &  &  & $3/2^+$ & 3/2 & $-\frac{1}{2}\sqrt{\frac{1}{5}}$ & 0 \\\cline{2-7}
 & $5/2(3)^+$ & 2 & $1/2^+$ & 1/2 & 0 & $-\frac{2}{3}\sqrt{\frac{2}{35}}$ \\
 &  &  & $3/2^+$ & 1/2 & 0 & $-\frac{2}{3}\sqrt{\frac{1}{35}}$ \\
 &  &  & $3/2^+$ & 3/2 & 0 & $\sqrt{\frac{2}{105}}$ \\\cline{2-7}
 & $7/2(3)^+$ & 2 & $1/2^+$ & 1/2 & 0 & $-\sqrt{\frac{2}{105}}$ \\
 &  &  & $3/2^+$ & 1/2 & 0 & $-\frac{1}{\sqrt{105}}$ \\
 &  &  & $3/2^+$ & 3/2 & 0 & $-\frac{1}{\sqrt{21}}$ \\
\end{tabular}
\end{center}
\end{table}

The amplitudes for the decays of the positive parity excitations with
$\lambda$-$\rho$ mixed excited states $(\ell_\lambda,\ell_\rho)=(1,1)$ 
of $\Lambda_c^*(J^-)\rightarrow \Sigma_c^{(*)}\pi$ 
are given by 
\begin{widetext}
\begin{equation}
 \begin{split}
  -iA_h^{\nabla\cdot\sigma} &= G
\dfrac{\omega_\pi}{m} 
\left\{\tilde{c}_1
  \left((-1)^\ell2a_\rho\dfrac{q_\lambda}{a_\lambda}
        +a_\lambda\dfrac{q_\rho}{a_\rho}
\right)
+\tilde{c}_3
\dfrac{q_\lambda q_\rho}{a_\lambda a_\rho}
  \left(\dfrac{1}{2}q_\lambda+q_\rho\right)
\right\}F(q),
\\
  -iA_h^{q\cdot\sigma} &= G
  \dfrac{q}{m}\left(\dfrac{M}{2m+M}\omega_\pi-2m\right)
(-1)\tilde{c}_3 \dfrac{q_\lambda q_\rho}{a_\lambda a_\rho}F(q),
 \end{split}
\label{eq:pp}
\end{equation}
\end{widetext}
{where $\ell$ denotes the total angular momentum
$\vec{\ell}=\vec{\ell}_\lambda+\vec{\ell}_\rho$ and} the coefficients $\tilde{c}_1$ and $\tilde{c}_3$ are summarized in
Table~\ref{tab:pp}.

\section{Matrix elements in the heavy quark limit}

In this appendix, we derive the matrix elements in the heavy quark limit
to show how and when the geometric factor is separated, leading to the 
model independent relations~\cite{Isgur:1991wq}.  
Let us consider one pion emission of a heavy baryon containing one heavy quark $Q$
and a pair of light quarks $qq$.  
Following the notation in this paper, let the initial baryon denoted by 
$\Lambda_{Q}$ and the final one by {$\Sigma_{{Q}}$}.  
Then the spin and angular momentum couplings for the initial $\Lambda_c$
and  final $\Sigma_c\pi $ states are
\be
|i\ket &=& |\Lambda_c\ket =  [ j_{\Lambda_{{Q}}}, s_{Q} ]^{J_{\Lambda_{{c}}} M_{\Lambda_{{Q}}}},
\nonumber \\
|f\ket &=& |\Sigma_c\pi \ket = [ Y_L, [ j_{\Sigma_{{Q}}}, s_{Q} ]^{J_{\Sigma_{{Q}}}}  ]^{J_f M_f},
\ee
where $J_{\Lambda_{{Q}}, \Sigma_{{Q}}}$ is the baryon spin, $j_{\Lambda_{{Q}}, \Sigma_{{Q}}}$ the brown muck (light degrees of freedom) 
total spin, $L$ the relative  angular momentum of $\pi \Sigma_{{Q}}$, 
and $J_f$ is the total spin $J_{\Sigma_{{Q}}} + L$.  
The decay probability is then computed as
\be
\Gamma \sim \sum_L \vert \bra f | \calL_{int}|i\ket \vert ^2,
\ee
where $\calL_{int}$ is the pion-quark interaction, 
and the sum over final state is taken over possible $L$'s.  
For instance, for the decay of $5/2^+ \to 3/2^+$, the angular momentum $L$ can be
both 1 ($P$-wave) and 3 ($F$-wave), while for the decay of $5/2^+ \to 1/2^+$, 
only $F$-wave  is possible.  

In the literature, the model independent relation has been discussed for the ratio
of the decays into $\Sigma^*_c(3/2^+)$ and into  $\Sigma_c(1/2^+)$.  
In the heavy quark limit it can be obtained only for  
the decay into the same and single partial wave $L$.  
As we have discussed in {\ref{sec:higher}} in detail, this is possible only 
in some limited cases where a selection rule due to the diquark transitions 
imposes an additional constraint.  
For a single $L$, after recouping the final state,  we  obtain the
matrix element as follows,  
\be
& & \bra [ Y_L, [ j_{\Sigma_{{Q}}}, s_{Q} ]^{J_{\Sigma_{{Q}}}}  ]^{J_f M_f} \vert
\calL_{int}
\vert [ j_{\Lambda_{{Q}}}, s_{Q} ]^{J_{\Lambda_{{Q}}} M_{\Lambda_{{Q}}}} \ket \nonumber \\
&=&
\sum_{j_f}
\hat J_{\Sigma_{{Q}}} \hat j_f (-1)^{j_{\Sigma_{{Q}}}+S_Q+J_f+L}
\left\{
\begin{array}{ccc}
J_{\Sigma_{{Q}}} & j_{\Sigma_{{Q}}} & s_Q \\
j_f & J_f & L
\end{array}
\right\} \nonumber\\
& &
\bra [ [Y_L,  j_{\Sigma_{{Q}}}]^{j_f}, s_{Q}]^{J_f M_f} \vert
\calL_{int}
\vert [ j_{\Lambda_{{Q}}}, s_{Q} ]^{J_{\Lambda_{{Q}}} M_{\Lambda_{{Q}}}} \ket \ .
\label{eq:HQ_ME}
\ee
Because the interaction $\calL_{int}$ is active only for the light quarks, 
after the application of the Wigner-Eckart theorem, 
the matrix element in the {third} line can be factorized into the one of 
the light degrees of freedom,  
$\bra  [Y_L,  j_{\Sigma_{{Q}}}]^{j_f} \vert\vert
\calL_{int}
\vert \vert  j_{\Lambda_{{Q}}} \ket$, 
and the trivial one of the heavy quark.
If, furthermore, the brown muck configuration is uniquely determined, 
which is to fix $j_f$  at a single value, 
the $J_{\Sigma_{{Q}}}$ dependence is completely dictated by the 
6-$j$ symbol and the normalization  $\hat J_{\Sigma_{{Q}}}$.  
This explains how and when the ratio in Eq.~(\ref{eq:R}) can be determined in a model independent manner
by the formula (\ref{eq:HQ_ME}).

\bibliography{charm.bib}
\end{document}